\documentclass[copyright,creativecommons]{eptcs}
\providecommand{\event}{HSB 2013}

\providecommand{\firstpage}{1}
\usepackage{graphics}
\usepackage{epsfig}
\usepackage{amsmath}
\usepackage{amssymb}
\usepackage{theorem}
\usepackage{algpseudocode}
\usepackage{algorithm}
\usepackage{subfig}
\newcommand{\STL}{STL*}

\begingroup\lccode`~=`*\lowercase{\endgroup
	\edef~{\noexpand\sp{\ast}{}}}
\mathcode`*="8000

\renewcommand{\*}[1]{\ensuremath{\sp{\ast_{#1}{}}}}

\makeatletter
\newcommand{\frz}{\@ifstar\@freeze\@ifreeze}
\newcommand{\@ifreeze}[1]{\ensuremath{\operatorname{\ast}_{#1}}}
\newcommand{\@freeze}{\ensuremath{\operatorname{\ast}}}

\newcommand{\@STLop}[2]{\ensuremath{\operatorname{\textbf{#1}}_{#2}}}
\newcommand{\@STLopNS}[2]{\@STLop{#1}{[#2]}}
\newcommand{\U}{\@ifstar{\@STLop{U}}{\@STLopNS{U}}}
\newcommand{\G}{\@ifstar{\@STLop{G}}{\@STLopNS{G}}}
\newcommand{\F}{\@ifstar{\@STLop{F}}{\@STLopNS{F}}}
\newcommand{\R}{\@ifstar{\@STLop{R}}{\@STLopNS{R}}}
\makeatother

\newcommand{\thelang}{\ensuremath{\mathcal{L}}}
\newcommand{\lang}[1][t,t*]{\ensuremath{\thelang_{#1}}}

\makeatletter
\newcommand{\@boldop}[1]{\ensuremath{\operatorname{\mathbf{#1}}}}
\newcommand{\dist}{\@boldop{dist}}
\newcommand{\depth}{\@boldop{depth}}
\newcommand{\Dist}{\@boldop{Dist}}
\makeatother

\newcommand{\II}{\mathcal{I}}

\newcommand{\store}[3][t*]{\ensuremath{#1[#2\leftarrow#3]}}
\title{Robustness Analysis for Value-Freezing Signal Temporal Logic}
\author{L.~Brim, T.~Vejpustek, D. \v{S}afr\'anek, and J.~Fabrikov\'a
\thanks{The work has been supported by the Grant Agency of Czech Republic grant GAP202/11/0312 and by the EC OP project No. CZ.1.07/2.3.00/20.0256.}
\institute{Faculty of Informatics\\
Masaryk University\\
Botanick\'a 68a, Brno, Czech Republic}
\email{safranek@fi.muni.cz}
}
\def\titlerunning{Robustness Analysis with STL*}
\def\authorrunning{L.~Brim et al.}

\newtheorem{theorem}{Theorem}[section]
\newtheorem{defn}{Definition}[section]

\newcommand{\dom}{\operatorname{\mathbf{dom}}}

\renewcommand{\phi}{\varphi}
\newcommand{\e}{\varepsilon}
\newcommand{\vc}[1]{\mathbf{#1}}
\newcommand{\RR}{\mathbb{R}}
\newcommand{\NN}{\mathbb{N}}

\newcommand{\authors}[1]{#1~et~al.}
\newcommand{\procedure}[1]{\textsc{#1}}

\renewcommand{\a}{\alpha}

\begin{document}
\maketitle

\begin{abstract}
In our previous work we have introduced the logic STL*, an extension of Signal Temporal Logic (STL) that allows value freezing. In this paper, we define robustness measures for STL* by adapting the robustness measures previously introduced for Metric Temporal Logic (MTL). Furthermore, we present an algorithm for STL* robustness computation, which is implemented in the tool Parasim. Application of STL* robustness analysis is demonstrated on case studies.
\end{abstract}

\section{Introduction}

% robustnost je dobry pristup k analyze spojitych systemu, zejmena vyuziti v systemove biologii, priklady pouziti (Robustness diskrimination...)

A~particular place among formalisms adopted by systems biology is occupied by temporal logics,
which serve as a~language for description of biological systems behaviour. Resulting temporal formulae can be
used during computer-aided system analysis, such as model checking \cite{modelchecking}, which automatically verifies whether
a~model satisfies given temporal formula. Methods based on temporal logics have been successfully employed to study
biological phenomena~\cite{STL-use,biocham-practice,eziocav11} (see~\cite{SFM} for review).

% logiky pro signaly (MTL, STL)

Since most of current models developed in computational systems biology have the form of ordinary differential equations, model checking cannot be directly employed and is typically replaced with a non-exhaustive procedure of monitoring~\cite{Maler_STL}. In this setting, a (finite) set of signals representing individual time-courses of the model is monitored wrt a given temporal specification. In particular, the respective temporal logics are interpreted over individual signals that are most typically simplified to discrete timed state sequences (time series) approximating the continuous trajectories by means of numerical simulation. Temporal logics fitting this interpretation are Metric Temporal Logic (MTL)~\cite{MTL} and Signal Temporal Logic (STL)~\cite{Maler_STL}, which allow quantifying modalities with the time frame represented by a closed time interval.
MTL possesses both discrete and continuous semantics, as it can be interpreted over both infinite timed state sequences and continuous signals. STL is practically focused and is defined for piece-wise linear approximations of continuous signals. 

% motivace pro value-freezing (opet biologie, oscilace)
Temporal logics are satisfactorily used in systems biology to express statements about a~single instance of system behaviour such as \emph{in five minutes, concentration
of glucose will be greater than 0.8}. However, many biological hypotheses contain relative temporal references, e.g., \emph{after protein $P$ reaches the maximum concentration, a steady concentration of $P$ is reached which is less than half of the maximum}. Such a scenario can be found, e.g., in feed-forward genetic regulatory circuits generating pulses in expression signals~\cite{kaplanthe2008}. In common temporal logics, such a general query cannot be expressed. This is because the values in different time points cannot be compared, i.e., the property \emph{in five minutes, concentration
of glucose will rise by 0.2}, which relates glucose concentration at current time and in the future, cannot be specified. Of specific interest is oscillatory behaviour, e.g., a sequence of gradually increasing peaks followed by a limit cycle with a stable amplitude~\cite{limitcycles}. In order to express the increasing amplitude, it is necessary to detect local extremes in signals and compare respective signal values. This cannot be achieved using common temporal logics. Signals with a series of increasing local maxima have been observed, e.g., in response of FGF signalling pathways transferring stimuli from mutated FGFR3 receptors to target effectors affecting bone cells growth~\cite{Krejci2004152}. Since the mentioned behaviour correlates with the phenotype of dysplasia, it is necessary to develop models that mechanistically capture the respective signalling pathways and to analyse circumstances under which the undesired behaviour occurs. This makes a necessary step before designing a targeted medical treatment. To this end, temporal logics and verification procedures which allow to capture and analyse such complex phenotypes have to be developed.        

In~\cite{Dluhos-stlstar},
we have introduced a~new temporal logic \STL{} which alleviates limitations mentioned above. Expressiveness of \STL{} is enhanced by signal-value
freeze operator which stores values at certain time, which may be referred to in the future. This allows \STL{} to specify and distinguish various dynamic aspects which occur in biological systems, in addition to the phenomena mentioned above, these can be, e.g., damped oscillations~\cite{sybi-periodic} or local extremes in species concentration. It is worth noting that some more complex queries can be expressed in traditional temporal logic by including signal derivatives into atomic propositions. However, this does not directly apply to queries mentioned above. One can express the presence and shape of a local extreme by using the first and second derivative, but still the values in particular time points have to be compared in order to express the complex queries.
%without the need to compute and reason about derivatives of analysed signals~\cite{Dluhos-stlstar}.

% robustnost
An important concept associated with biological systems and temporal logics is \emph{robustness}, the ability of a~system 
to maintain its function against perturbations \cite{Kitano-robustness}. Since system function can be expressed in the terms
of temporal logic, we speak of robustness with respect to a~temporal logic formula, which can be quantified and
computed~\cite{FP-robustness,robustness-property}. Robustness significantly enhances model analysis and gives an optimization goal for model parameter estimation/synthesis~\cite{STL-robustness,STL-parameters,FagesRobustness}.

% nas prinos

This paper %contributes by bringing
introduces the notion of robustness in the value-freezing logic \STL{} setting. In particular, we extend the continuous and discrete measure defined for MTL by \authors{Fainekos} \cite{FP-robustness} to the semantic domain of \STL{}. Robustness of the input signal with respect to \STL{} formula delineates the robust neighbourhood of the signal (the maximal ``tube'' around the signal where the formula is satisfied). The robustness measure we propose (Section~\ref{sec:robustness}) is defined inductively wrt the formula structure and is based on a distance metrics employed on the signal domain extended with (multiple) dimensions representing the frozen time points. The theoretical framework is computationally supported with an algorithm based on solving the optimization problem (Section~\ref{sec:computation}) provided that the logic is restricted to linear predicates. Special consideration is given to optimization of the formula to overcome unnecessary computational overhead. 

Implementation of our algorithm is included as a part of Parasim~\cite{parasim}, a~tool aimed as a modular environment for monitoring and robustness analysis of kinetic models. To demonstrate the usage and evaluate the performance, we present case studies of two simple kinetic models (Section~\ref{sec:casestudy}). 

%\enlargethispage{8mm}

\vspace*{-3mm}

\subsection{Related Work}
% STL (Maler et al.), logika a monitoring + STL* -- HSB 2012
% zminit freeze (viz DP 4.0), zbytecne...
% koncept robustnosti existuje pro MTL a STL, a je vhodne rozsirit i na value-freezing logiky -- srovnat Donze, Fainekos, BioCHam, CAV 2013???

Robustness measures have been defined for three temporal logics targeting deterministic continuous systems: STL~\cite{STL-robustness}, MTL~\cite{FP-robustness} and QFLTL~\cite{robustness-property}. We adopt the concept of behaviour-based robustness introduced on a fragment of MTL by Fainekos et al.~\cite{FP-robustness}, who define robustness measure for MTL formulae with discrete \cite{FP-robustness0} and continuous \cite{FP-robustness} semantics. In \cite{FP-robustness}, Fainekos et. al prove a~theorem connecting discrete and continuous robustness, which is valuable for robustness computation. A~recent tool \cite{FP-taliro} implements the method. \authors{Donz\'{e}}~\cite{Maler_STL} use STL to define a~distinct robustness measure, albeit constructed from~\cite{FP-robustness}, and propose its application for space exploration \cite{STL-robustness,STL-parameters}, which was implemented in the Breach Toolbox~\cite{STL-breach}. The work is further improved from the computational point of view in~\cite{DFM13}. Our implementation (Parasim) is based on a simplified version of the robustness analysis algorithm for STL where the sensitivity-based computation of local robustness is replaced with direct computation of trajectories distance. The extension for \STL{} as presented in Section~\ref{sec:computation} is implemented in this setting. 

Fages et al.~\cite{robustness-property} introduced property-based approach to robustness that fixes input behaviour and examines the formula. Basically, it measures the extent to which the formula can be modified while preserving its satisfaction. The tool BioCham implements this idea~\cite{biocham}. Extended LTL logic with constraints over real numbers (quantifier-free LTL) is employed being defined for finite discrete time-series.

It is worth noting that the problem of formula satisfiability is undecidable for MTL \cite{MTL}. To achieve decidability, Alur and Henzinger specified further conditions on intervals associated with temporal operators~\cite{MITL}. The result, metric interval temporal logic, requires all intervals to be non-singular and is interpreted over timed state-sequences where time points are replaced with consecutive time intervals. STL was introduced by Maler and Nickovic in \cite{Maler_STL} as a~basis for their monitoring procedure. Technically, it comprises a~variant of MITL
interpreted over real signals. Because of its practical purpose, in~\cite{Dluhos-stlstar} we selected STL as a good candidate for extension with value-freezing.

\section{Background}

\STL{} is evaluated over finite time continuous signals (finite signals for short).

\begin{defn}\label{def:signal}
	%Let $X$ be the domain of states of a~system
	Let $n\in\NN$ and $T=[0,r]$ where $r\in\RR^+$. Then $s:T\to\RR^n$ is a~\emph{bounded continuous-time signal} and $T$ its \emph{time domain}.
	We denote $l(s)=r$ the length of signal~$s$.
\end{defn}

Signal value freezing is facilitated by the following structure which is used to store time values at various time points which then can be referred to in predicates.

\begin{defn}\label{def:frozen_time}
	Let $\II$ be a~finite index set. \emph{Frozen time vector} is a~function:
	\begin{equation*}t*:\II\to\RR^+_0\end{equation*}
\end{defn}
%Restriction to finite number of frozen times evidently does not reduce logic expressiveness as formulae are finite structures
%and thus cannot contain infinite number of freeze operators or infinite predicates.
The symbol $t_i*=t*(i)$ is referred to as $i$-th frozen time. For convenience reasons and without
loss of generality, we will henceforth assume that an index set $\II=\{1,\ldots,m\}$ is given, where $m\in\NN$.

Predicates comprise Boolean expressions over values of a~signal $s$ at time $t$ and each frozen time $t_i*$, where $x_j$~denotes the $j$-the component of the signal at time~$t$,
i.e. $s(t)=(x_1,\ldots,x_j,\ldots,x_n),$ and $x_j\*i$ the $j$-th component at time~$t_i*$. When $|\II|=1$, we usually omit the index of asterisk, e.g. $x_i*=x_i\*1$.

We consider only predicates given by linear inequalities, so that analytic expressions of predicate robustness is possible.

%Predicates over real signals can now be defined as subsets of $\RR^n\times\left(\RR^n\right)^\II$. Only predicates given by linear inequalities are considered though,
%which enables analytical expression of predicate robustness.

%For time point $t$, frozen time vector $t*$ and a~$n$-dimensional signal $s$, 

\begin{defn}\label{def:stl*predicate}
	Let $n\in\NN$, $b\in\RR$ and $a_{ij}\in\RR$ where $i\in \{0\}\cup\II$, $j\in\{1,\ldots,n\}$ and not all $a_{ij}$ are zero. A~\emph{predicate} is defined as a~subset
	of $\RR^n\times\left(\RR^n\right)^\II$ such that:
	\begin{equation*}\sum_{j=1}^na_{0j}x_j+\sum_{i=1}^{|\II|}\sum_{j=1}^na_{ij}x_j\*i+b\ge 0\end{equation*}
\end{defn}

Predicates are specified by the set of associated coefficients $a_{ij},b$ (where coefficients $a_{0j}$ are connected with the current time $t$). Therefore,
for convenience reasons, we will use these coefficients to represent predicates. Predicates with all coefficients $a_{ij}$ zero were omitted since they are
of the form $b\ge 0$ and, therefore, trivially true or false.%$\top$ or $\neg\top$ can be used in their stead.

%Predicates are uniquely determined by the set of associated coefficients $a_{ij},b$ .
%Therefore, for convenience reasons, we will use these coefficients to specify predicates.

Predicates with equality (i.e. having $=$ in place of $\ge$), although theoretically possible, lack practical value, as they are not robust (small perturbation may invalidate the property).
This has been already argued in \cite{Dluhos-stlstar}, albeit without defining the concept of robustness. Since robustness of predicates with strict and non-strict inequalities does not differ, we consider only non-strict inequalities.

Freeze operator is used to store the time point into frozen time vector, thus facilitating signal value freezing. The following definition introduces an auxiliary concept of storing the current time $t$ as the $i$th component of the frozen time vector.
\begin{defn}\label{def:stl*store}
	Let $t*$ be frozen time vector, $i,j\in\II$ and $t\in\RR^+_0$. \emph{Freezing $i$th component of $t*$ in $t$} is
	denoted as $\store{i}{t}$ and defined:
	\begin{equation*}
		\store{i}{t}(j)=\begin{cases}
			t	&i=j\\
			t_j*&i\neq j\\
		\end{cases}
	\end{equation*}
\end{defn}

\begin{defn}\label{def:stl*syntax}
	Syntax of \STL{} is defined by the following grammar:
\begin{equation*}\phi::= \mu \mid \top \mid \neg\phi \mid \phi_1\lor\phi_2 \mid \phi_1\U*I\phi_2 \mid \frz i\phi\end{equation*}
	where $i\in\II$, $\top$ denotes the true constant, $\mu$ is a~predicate as of Definition~\ref{def:stl*predicate}
	and $I\subseteq\RR^+_0$ a~closed non-singular interval.
\end{defn}
Note that all Boolean connectives and temporal operators $\F*{}$ and $\G*{}$
can be defined using the basic operators defined above.
Similarly to predicates, when $|\II|=1$, we usually omit the index of freeze operator, as in $\frz*\G*I(x>x*)=\frz1\G*I(x>x\*1)$.
Henceforth, let $i,\mu,\phi,\phi_1,\phi_2$ be the same as in Definition~\ref{def:stl*syntax}.
%The further text will consider only \STL{} formulae.

\begin{defn}\label{def:stl*semantics} Let $s\in\left(\RR^n\right)^T$ be a~signal, $t\in T$ a~time point and $t*\in T^\II$ a~frozen time vector.
	Formula satisfaction is defined inductively:
\begin{equation*}\begin{array}{lcl}
(s,t,t*)\models\top					&& \\
(s,t,t*)\models\mu				&\iff& (s(t),s \circ t*)\in\mu\\
(s,t,t*)\models\neg\phi			&\iff& (s,t,t*)\not\models\phi\\
(s,t,t*)\models\phi_1\lor\phi_2	&\iff& (s,t,t*)\models\phi_1\lor(s,t,t*)\models\phi_2\\
(s,t,t*)\models\phi_1\U*I\phi_2	&\iff& \exists\ t'\in t\oplus I:(s,t',t*)\models\phi_2\,\land\\
									&& \forall\ t''\in[t,t']:(s,t'',t*)\models\phi_1\\
(s,t,t*)\models\frz i\phi		&\iff&(s,t,\store{i}{t})\models\phi
\end{array}\end{equation*}
\end{defn}
Operator $\circ$ is used to denote function composition, i.e. $(s\circ t*)\in\left(\RR^n\right)^\II$ and $(s\circ t*)(i)=s(t_i*)$
and $t\oplus I$ stands for $\{t+u\mid u \in I\}$.

\begin{defn}Let $s\in\left(\RR^n\right)^T$ be signal and $\phi$ formula. Formula satisfaction by signal is given:
	\begin{equation*}s\models\phi \iff (s,0,\vc{0})\models\phi\end{equation*}
	where $\vc{0}$ denotes the \emph{zero frozen time vector}, i.e. $\{(i,0)|i\in\II\}$.
\end{defn}

Intuitively, interpretation of $\frz i\phi$ is the following: freeze operator stores signal values at the time of $\frz i\phi$ evaluation,
which can then be referred to using index $i$ in predicates of $\phi$. An example property, \emph{``in the next five time units,
$x$ increases by $8$''} can be specified as:
\begin{equation*}\frz*\F{0,5}(x\ge x*+8)\end{equation*}
where $x*$ refers to value of $x$ at time $0$.

When intervals associated with until operators are bounded, satisfaction of a given formula can be decided on any finite signal of sufficient length.
This length can be determined from the formula structure in a~way similar to \cite{Maler_STL} and corresponds to the furthest time point (among all possible signals) which
has to be examined in order to determine formula satisfaction. This clearly also holds for frozen time values.

%Similarly to \cite{Maler_STL}, necessary input length can be defined, which determines the furthest time point which needs to be examined in order to determine whether
%$s\models\phi$. This clearly also holds for frozen time values.%It is not affected by freeze operator, as shown by the following theorem.

%\begin{theorem}\label{thm:tstar_cut-continuous} % tahle celá část se pravděpodobně vynechá
%	Consider arbitrary formula $\phi$, signal~$s$ and $i\in\II$. Then $t\ge t_i*$ in all expressions in the form $(s,t,t*)\models\psi$ obtained when expanding $s\models\phi$
%	inductively to formula structure according to Definition~\ref{def:stl*semantics}.
%\end{theorem}
%\begin{proof}
%	Inductive to the expansion of $s\models\phi$. The base case, $(s,0,\vc{0})\models\phi$, is obvious. As rules for predicates, $\top$, $\neg$ and $\lor$ do not
%	change $t$ and $t*$, we need only consider until and freeze operators.

%	Let $t\ge t_i$ and consider expression $(s,t,t*)\models\phi_1\U*I\phi_2$. Its expansion contains $(s,t',t*)\models\phi_2$ and $(s,t'',t*)\models\phi_2$
%	where $t'\in t\oplus I$ and $t''\in[t,t']$. Obviously both $t'>t\ge t_i*$ and $t''>t\ge t_i*$.

%	Let $t\ge t_i$ and consider expression $(s,t,t*)\models\frz i\phi$, which is equal to $(s,t,\store{i}{t})\models\phi$. By definition, $\store{i}{t}_i=t\le t$.
%\end{proof}
%This theorem (or rather its discrete analogue Theorem~\ref{thm:tstar_cut-discrete}) also ensures that monitoring algorithm need not analyse $t*>t$.

\begin{defn}\label{def:necessary_length}
	Let $\phi$ be a~formula. The \emph{necessary input length} for $\phi$, $l(\phi)$ is defined inductively:
	\begin{align*}
		l(\top)=l(\mu)				&= 0\\
		l(\neg\phi)=l(\frz i\phi)	&= l(\phi)\\
		l(\phi_1\lor\phi_2)			&= \max(l(\phi_1),l(\phi_2))\\
		l(\phi_1\U*I\phi_2)			&= \max(l(\phi_1),l(\phi_2))+\sup I
	\end{align*}
\end{defn}
When $l(s)<l(\phi)$ we state that $s\not\models\phi$.

%Without loss of generality all input behaviour will be assumed of at least necessary length. Formally:
%\begin{defn}
%	Let $\mu$ be a~predicate and $s$ signal. It is assumed $(s,t,t*)\not\models\mu$ if $t>l(s)$, or $t_i*>l(s)$ for any $i\in\II$.
%\end{defn}

%má se někde zavádět sémantická ekvivalence ?
%For convenience, we define the concept of semantic equivalence: %definice?
%\begin{equation*}\phi\sim\psi\iff \left(\forall s,t,t*:(s,t,t*)\models\phi\iff(s,t,t*)\models\psi\right)\end{equation*}

%tady by možná mohl být podnadpis -- a nebo by se to mohlo dát někam dopředu

Frozen time indices and freeze operators share some similarities with variables and quantifiers of predicate logic. We may distinguish free and
bound indices, where index $i$ is free if it is used in a~predicate (i.e. coefficient $a_{ij}$ is not zero for some $j$) and is not in the scope of 
operator $\frz i$.

%\begin{enumerate}
%	\item Index $i$ is free in a~predicate if some coefficient $a_{ij}$ is not zero, i.e. when $i$ is used in the predicate.
%	\item Boolean and temporal logical connectives correspond to union of subformulae free indices.
%	\item Operator $\frz i$ removes $i$ from the free indices.
%\end{enumerate}
Naturally, whenever $i$ is free in $\phi$, then $s\models\phi$ iff $s\models\frz i\phi,$ since $t_i*$ is zero in both cases.

Additionally, we may substitute for free indices of a~formula in a~manner similar to variable substitution. However, it only makes sense to substitute one index for another,
which we will denote \emph{index renaming} and express as $\phi[\pi]$ where $\pi$ is a~total function on $\II$ (but not necessarily a~permutation -- two indices can be renamed to one) or $\phi[k/l]$,
where $k$ is renamed to~$l$. To preserve formula semantics, renaming is only safe when no free index becomes bound after renaming in any subformula.

% syntaxe a semantika STL* 

% definovat spojitou semantiku a sdeli ze diskretni lze ze spojite, pokud je diskrerni nutna, ale tu bychom nejradeji vynechali -- vynechame, protoze mame inzen., pristup

% volny a vazany indexy  

\section{Robustness Measures for STL*}
\label{sec:robustness}

Following from \STL{} semantics, robustness of signal~$s$ with respect to formula~$\phi$ is given for each time point $t$ and frozen time vector $t*$ and denoted by $\rho(\phi,s,t,t*)$.
We also define $\rho(\phi,s)=\rho(\phi,s,0,\vc{0})$. Robustness of signal $s$ with respect to formula $\phi$ is a~value, which under-approximates the distance of $s$ from the set
of signals where $\phi$ has different truth value~\cite{FP-robustness}. To express this formally, we first need to define certain basic concepts (where $S$ is a~set of signals):
\begin{itemize}
	\item Distance of signals is given by their maximum pointwise distance: $d(s,s')=\max_{t\in\RR^+_0}d(s(t),s'(t))$
	\item \emph{Set distance} is given by minimum distance to the set: $\dist(s,S)=\min\{d(s,s')\mid s'\in S\}$
	\item \emph{Set depth} is given by set distance to the complement: $\depth(s,S)=\dist\left(s,\overline{S}\right)$
	\item \emph{Signed distance} is given: $\Dist(s,S)=\begin{cases}-\dist(s,S)&s\notin S\\\depth(s,S)&s\in S\end{cases}$
\end{itemize}

\begin{figure}[h!]
	\centering
	\includegraphics[width=0.4\textwidth]{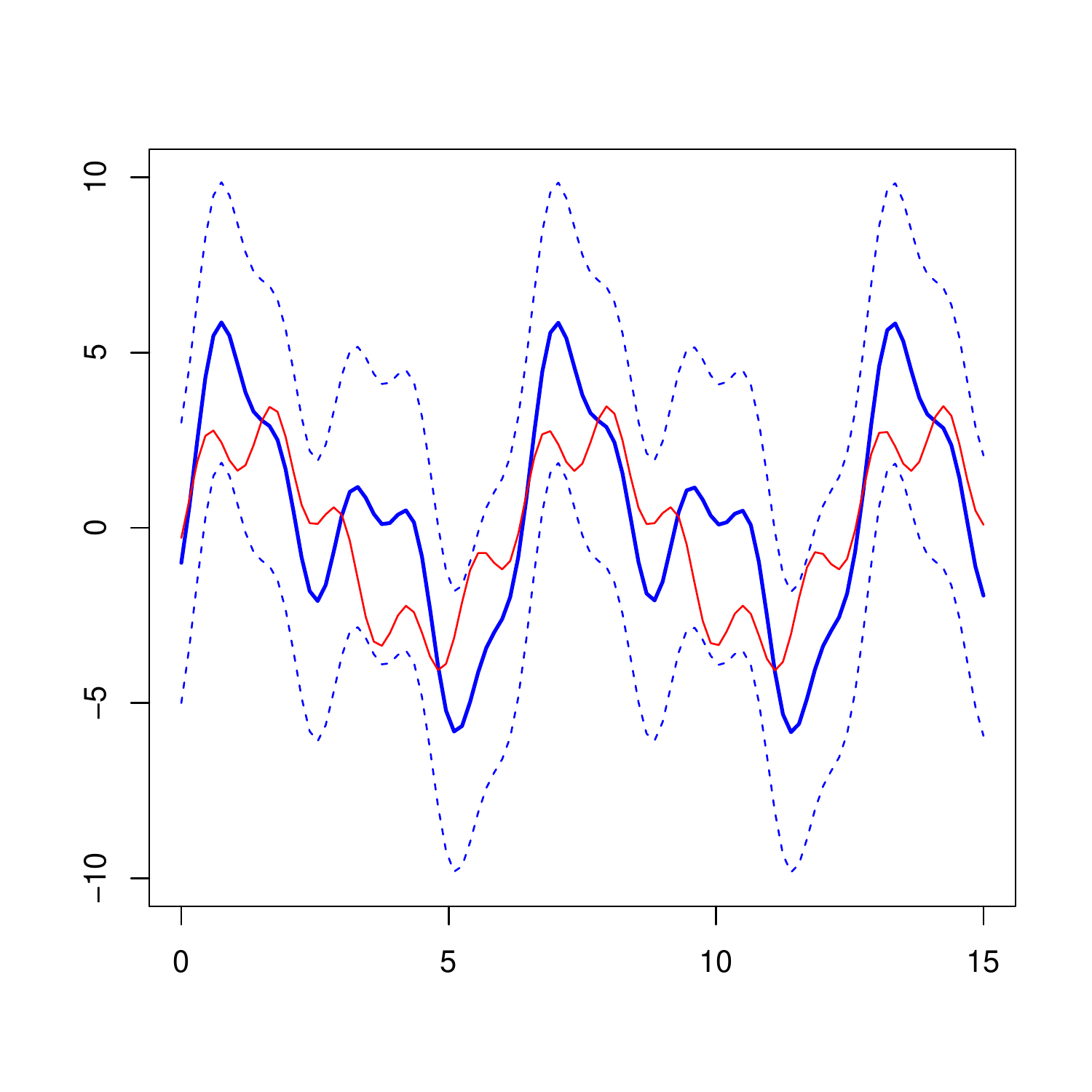}
	\caption[Signal and robust neighbourhood.]{Signal~$s$ (blue, thick) and borders of its robust neighbourhood (blue, dashed) with an example of a~signal (red) contained in the robust neighbourhood (adapted from \cite{FP-robustness}).}\label{fig:tube}
\end{figure}

The value $\rho(\phi,s)$ underapproximates the signed distance of $s$ from the set of all signals satisfying~$\phi$, $\thelang(\phi)$, i.e.
$|\rho(\phi,s)|\le|\Dist(s,\thelang(\phi))|$ holds while their signs are identical. The absolute value of $\rho(\phi,s)$ thus delineates
an equidistant tube where all signals satisfy $\phi$ if and only if $s$ does -- the \emph{robust neighbourhood} of~$s$ (see Figure~\ref{fig:tube}).

It would be desirable to define the robustness equal to the signed distance; however, by~\cite{FP-robustness}, the robustness computation would not be feasible then. In order to be sound, the robustness definition has to satisfy the following property (for any $\phi$, $s$, $t$ and $t*$):
\begin{equation}\label{eq:robust_sound}
	-\dist\left(s,\lang(\phi)\right)\le\rho(\phi,s,t,t*)\le\depth\left(s,\lang(\phi)\right),
\end{equation}
where $\lang(\phi)=\{s\mid (s,t,t*)\models\phi\}$. Since $\depth(s,\lang(\phi))=0$ when $(s,t,t*)\not\models\phi$ (and analogously for $\dist$), this actually requires that:
\begin{enumerate}
	\item $s\models\phi\implies 0\le\rho(\phi,s,t,t*)\le\depth\left(s,\lang(\phi)\right)$,
	\item $s\not\models\phi\implies -\dist\left(s,\lang(\phi)\right)\le\rho(\phi,s,t,t*)\le 0$.
\end{enumerate}

Robustness is defined inductively for each logical connective from its semantics in such manner that Boolean functions $\land$ and $\lor$ are replaced by real functions $\min$ and $\max$ (respectively).
Quantifiers in the semantics of operator $\U*{}$ can then be expressed by infinite disjunction or conjunction. Robustness wrt predicate $\mu$ is defined as $\Dist(s,\lang(\mu))$, i.e. the ideal value
without underapproximation. If $\rho(\mu,s)$ was lower, it would diminish resulting robustness value, for robustness wrt formula cannot be greater than robustness wrt any of its predicates.
Soundness of this definition (property \eqref{eq:robust_sound}) is, naturally, proved inductively wrt formula structure.

This has already been established by \authors{Fainekos} in~\cite{FP-robustness}, albeit for MTL which does not allow signal value freezing. Nevertheless,
their definition can be directly extended for \STL{}. Intuitively, this is due to frozen time values being only stored by freeze operators and retrieved in predicates,
which does not affect other logical connectives. The full proof can be found in \cite{src} (page 83).

Consequently, we have to define robustness for the freeze operator. It follows from its semantics:
\begin{equation*}\label{eq:op:frz1}
	\lang(\frz i\phi)=\{s \mid (s,t,t*)\models\frz i\phi\}=\{s \mid (s,t,\store{i}{t})\models\phi\}=\lang[t,\store{i}{t}](\phi)
\end{equation*}
Thus, robustness of freeze operator can be defined in the following manner:
\begin{equation*}\label{eq:op:frz2}
	\rho(\frz i\phi,s,t,t*)=\rho(\phi,s,t,\store{i}{t})
\end{equation*}
Assume $-\dist(s,\lang(\phi))\le\rho(\phi,s,t,t*)\le\depth(s,\lang(\phi))$ for any $t,t*$.
Therefore, it also holds for $t$ and $\store{i}{t}$ and thus:
\begin{equation*}
	-\dist(s,\lang[t,\store{i}{t}](\phi))\le\rho(\phi,s,t,\store{i}{t})\le\depth(s,\lang[t,\store{i}{t}](\phi))
\end{equation*}
From which follows the validity of \eqref{eq:robust_sound} for $\rho(\frz i\phi,s,t,t*)$. \STL{} robustness for logical connectives is presented in Figure~\ref{fig:robustness}.

\begin{figure}[h!]
	\centering
	\begin{equation*}\begin{array}{l@{\ =\ }l}
		\rho(\top,s,t,t*)				& +\infty\\
		\rho(\neg\phi,s,t,t*)			& -\rho(\phi,s,t,t*)\\
		\rho(\phi_1\lor\phi_2,s,t,t*)	& \max\left(\rho(\phi_1,s,t,t*),\rho(\phi_2,s,t,t*)\right)\\
		\rho(\phi_1\U*I\phi_2,s,t,t*)	& \max\limits_{t'\in t\oplus I}\min\left(\rho(\phi_2,s,t',t*),\min\limits_{t''\in[t,t']}\rho(\phi_1,s,t'',t*)\right)\\
		\rho(\frz i\phi,s,t,t*)			& \rho(\phi,s,t,\store{i}{t})
	\end{array}\end{equation*}
	\caption{Robustness of \STL{} logical connectives.}\label{fig:robustness}
\end{figure}

\subsection{Robustness of Predicates}

Finding $\Dist(s,\lang(\mu))$ generally constitutes a~convex analysis problem \cite{FP-robustness}. Thus, it could be solved using convex programming for each $t$ and $t*$,
which would, however, greatly increase computation time, and therefore, analytic solution is preferable. To this end, we have restricted \STL{} predicates to be linear.

For predicate $\mu$ with coefficients $a_{ij},b$, the problem of finding $\Dist(s,\lang(\mu))$ can be reduced to optimization of $f(\vc{d})=\max_i\sum_jd_{ij}^2$ (where $i\in\II$ and $j\in\{1,\ldots,n\}$) under the constraint
$\sum_i\sum_ja_{ij}d_{ij}+\e=0$ for some positive $\e$. This is a~non-trivial problem, since %the derivative of $f$ is not continuous.
$f$ is not differentiable at point $\vc{d}$ where $f(\vc{d})=\sum_jd_{kj}^2=\sum_jd_{lj}^2$ for some $k\neq l$.
To solve it, generalized method of Lagrange multipliers
from \cite{nonsmooth_analysis} was used, resulting in the following definition of the robustness $\rho$ (detailed derivation can be found in \cite{src} (page 47)).

\begin{defn}\label{def:predicate_robustness}
	Let $\mu$ be a~predicate with coefficients $a_{ij},b$. Then
	$$\rho(\mu,s,t,t*)=\frac{\sum_ja_{0j}s_j(t)+\sum_i\sum_ja_{ij}s_j\left(t_i*\right)+b}{\sum_i\sqrt{\sum_ja_{ij}^2}}$$
	for arbitrary $s$, $t$, $t*$, $i$ ranging over $\II$, $j$ ranging over $\{1,\ldots,n\}$.
\end{defn}
The numerator corresponds to the left-hand side value of the predicate. %For predicates where all coefficients $a_{ij}$ are zero,
%this formula is, obviously, invalid. However, since such predicates are of form $b\ge 0$, they can be replaced by $\top$ or $\neg\top$ before robustness computation.

It holds that $\rho(\mu,s,t,t*)=\Dist(s,\lang(\mu))$, unless some time points given by $t$ and $t*$ are equal. This originates from the optimization problem,
where $t_k*=t_l*$ (or $t=t_k*$) would constitute another constraint, which might change the solution.

Suppose that $t_k*=t_l*$ (reasoning for $t=t_k*$ is similar). We can merge (sum) coefficients $a_{kj}$ and $a_{lj}$ for any given $j$, which effectively reduces 
the number of considered frozen times. Robustness of predicates with merged coefficients is greater, since the denominator of definition~\ref{def:predicate_robustness} becomes smaller
as $\sqrt{\sum_j\left(a_{kj}+a_{lj}\right)^2}\le\sqrt{\sum_ja_{kj}^2}+\sqrt{\sum_ja_{lj}^2}$ due to triangle inequality. % holds.
Therefore, even if we disregard possible time point equality, property \eqref{eq:robust_sound} still holds. However, the greater the value of $\rho(\mu,s,t,t*)$ is,
the better approximation of $\Dist(s,\lang(\phi))$ is obtained. Therefore, we will investigate two distinct cases when time points can be equal:

%Suppose $t_k*=t_l*$ for some $k,l\in\II$ (when $t=t_k*$, the reasoning is similar). Then coefficients $a_{kj}$ and $a_{lj}$ for any given $j$ can be merged,
%resulting in a~predicate with coefficients $a'_{kj}=a_{kj}+a_{lj}$, $a'_{lj}=0$ and $a'_{ij}=a_{ij}$ for $i\notin\{k,l\}$. Now, the number of considered
%frozen times is effectively reduced as the adjusted predicate left-hand side value does not depend on $t_l*$. This procedure can be repeated until all
%coefficients corresponding to equal time points are merged. Now the formula of Definition~\ref{def:predicate_robustness} can be used.

%It should be noted, that robustness with merged coefficients is greater than robustness without them due to triangle inequality. Thus, even if we disregard equal time points,
%the robustness still underapproximates robust neighbourhood of a~signal. However, 

%Now, we may distinguish two cases of time points being equal:
\begin{enumerate}
	\item It happens consistently for given formula $\phi$ and predicate $\mu$, i.e. $\phi$ is built in such way that the same time value is stored by freeze operator associated with both indices, such as:
		\begin{equation*}\psi=\G*{I_1}(\frz i\neg \frz j\F*{I_2}(x\*i+x\*j\ge x))\end{equation*}
	\item It is a~result of $\phi\equiv\frz i\left(\phi_1\U*I\phi_2\right)$ (or similar formula) evaluation:
		\begin{equation*}\begin{split}
			(s,t,t*)\models\phi	&\iff(s,t,\store{i}{t})\models\phi_1\U{a,b}\phi_2\iff\\
								&\exists t'\in[a+t,b+t]:(s,t',\store{i}{t})\models\phi_2\land\forall t''\in[t,t']:(s,t'',\store{i}{t})\models\phi_1
		\end{split}\end{equation*}
		When $a=0$, it may occur that $t'=t$. Additionally, $t''\in[t,t']$, therefore, satisfaction of $\phi_1$ by $(s,t,\store{i}{t})$ has to be evaluated.
		The equality of $t$ and $i$-th frozen time may be propagated to predicates. We have decided to omit this case in order to simplify robustness computation.
\end{enumerate}

%While we have decided to omit the latter case to simplify robustness computation, the first case raises a~question
%whether such formula can be modified so that this kind of time point equality does not occur. Indeed, we have been able to formulate several formula equalities involving freeze operator which make
%this possible (their proof is included in \cite{src}):
%
\subsection{Improving Approximation}\label{sec:merge_indices}
The formula $\psi$ (see above) is obviously badly written, since it can be reformulated with only one frozen time index: $\G*{I_1}(\neg\frz*\F*{I_2}(2x*\ge x))$. This eliminates time point equality
and thus improves robustness approximation. We have formulated three rules which can be used to automatically rewrite formula so that it does not induce consistent time point equality
(while preserving its meaning):
\begin{enumerate}
	\item Freeze operator is distributive over Boolean connectives. Consequently, freeze operators can be moved down along the formula syntax tree until they reach a~temporal operator,
		predicate or another freeze operator. %, resulting in a~semantically equivalent formula where every temporal operator and predicate is preceded by a~chain of freeze operators.
	\item Freeze operator preceding predicate can be merged with the predicate (associated coefficients being merged with coefficients for unfrozen time).
	\item\label{rule:freeze-merge} Two consecutive freeze operators and their associated indices can be merged. However, in order to preserve the formula meaning,
		a~completely new index has to be chosen as the result of merging.
\end{enumerate}
Subsequently, all \STL{} formulae can be written in such manner that each freeze operator is followed by until operator,
which also ensures that all frozen time indices generally refer to distinct time points. Indeed, all meaningful
formulae (i.e. not serving to illustrate semantic peculiarities) in \cite{Dluhos-stlstar} are specified in this manner.

This reinforces the connection between temporal operators and freeze operators expressiveness. Subsequently, it may be practical
to define an alternate \STL{} syntax, where signal value freezing is directly tied to the until operator, such as $\phi_1\U*I\*i\phi_2\equiv\frz i\left(\phi_1\U*I\phi_2\right)$.
However, we do not deem it necessary, seeing that it entails no expressiveness gain. Moreover, the current syntax of \STL{} may permit shorter and more transparent formulae. %This is illustrated by the following example:
%\begin{align*}
%	\phi &=\G*{I_3}\frz i\left(\F*{I_1}\left(x\*i>x\right)\lor\F*{I_2}\left(x\*i<x\right)\right)\\
%	\psi &=\G*{I_3}\left(\frz i\F*{I_1}\left(x\*i>x\right)\right)\lor\left(\frz i\F*{I_2}\left(x\*i<x\right)\right)
%\end{align*}
%While obviously $\phi\feq\psi$, the fact that $x\*i$ in both predicates refers to the same value is more apparent in $\phi$.

It should be noted that although application of previous rules may increase number of indices used in a~formula (due to the rule \eqref{rule:freeze-merge} which introduces one new index), it does not increase
the number of free indices in each subformula. On the contrary, the number of free indices may decrease.

% definice (udelat kompaktni variantu -- vztahnout na prvoradovou logiku)

% interpretace jako podaproximace konv. okoli, reseni predikatu (rovnost dist s robustnosti), argmentace freeze operatoru

% komentovat to jako slozitou ulohu z konvexni analyzy, kterou jsme do detailu popsali v thesis

% theorem 4.7, 4.8, 4.9 (dukazy do apendixu)

\section{Computation}
\label{sec:computation}

%To compute (or monitor) robustness of continuous signals, two approaches are used, both of which sample the input signal into a~finite number of points.
%Whereas \authors{Fainekos} then employ discrete robustness semantics \cite{Fainekos}, i.e. compute robustness only for sampled points, \authors{Donz\'{e}} connect resulting points
%to form a~piecewise linear signal (which approximates the original signal) and use continuous robustness semantics \cite{Donze_Robustness}. The latter approach, however, may necessitate introduction
%of new points, %into the timed state sequence,
%and so requires a~more involved algorithm.

%The monitoring procedure defined by \authors{Dluho\v{s}}, which decides satisfaction of \STL{} formulae \citeneed, %(without multiple freeze indices),
%extends the approach of \authors{Donz\'{e}}, therefore, it would seem only natural to extend it further into an algorithm for robustness computation. However, it should be also noted
%that the implementation of this method was found inefficient \citeneed due to time-consuming operations over polygons.

To compute (or monitor) robustness of continuous signal, we use the approach of \authors{Fainekos} \cite{FP-robustness}, which is based on discrete robustness semantics.
%Rather than designing a~complex algorithm which may eventually prove impractical due to its execution time, we have decided to base robustness computation on
%discrete robustness semantics, while presuming the sampling of input signal to be sufficiently dense.
The following procedure is used:
\begin{enumerate}
	\item Sample input signal $s:T\to\RR^m$ into a~\emph{timed state sequence} $(\tau,\sigma):\NN\to T\times\RR^m$.
	\item Compute robustness over points of the resulting timed state sequence (i.e. the discrete robustness).
\end{enumerate}
This only approximates continuous robustness of $s$. When MTL robustness is concerned, \authors{Fainekos} give bound for error introduced by this approximation under certain conditions,
which can be summarized as signal sampling being sufficiently dense with respect to given formula. We assume this strong theorem translates to \STL{} (as \STL{} robustness extends MTL robustness)
and deem the previous procedure good approximation for an input signal with large enough sampling rate.

%Subsequently, it is assumed that this procedure constitutes an acceptable approximation of continuous robustness measure which is supported by results of \authors{Fainekos} \cite{Fainekos},
%although error introduced by such approximation has not yet been evaluated for \STL{} robustness measure.
%although it was only implied for \STL{} robustness measure (see \sectionname~\ref{sec:5:discrete}).

Before the robustness monitoring algorithm is described, we should note that it can also be used to decide formula satisfaction, since positive robustness implies formula satisfaction
(and negative its invalidity). However, when $\rho(\phi,s)=0$ no information about formula satisfaction can be derived. Additionally, robustness measure only underapproximates
the robust neighbourhood, and so the robustness value may be zero even if clearly $s$ satisfies $\phi$. Consequently, classical monitoring may produce more precise results.

Algorithm~\ref{alg:monitor}
%presented in this section
computes robustness for a~\STL{} formula and sufficiently long timed state sequence (which may constitute a~sampled signal). It copies inductive definition of robustness
%(see \appendixname~\ref{app:log:stlstar:robustness})
with recursive calls of procedure \procedure{Monitor} (line~\ref{alg:mon:monitor}), which computes robustness only in the points of given state sequence. Therefore,
instead of frozen time vector $t*:\left(\RR^+_0\right)^\II$, \emph{frozen state vector} $\iota*:\NN^\II$ is used. The computation starts at zero index and zero frozen state vector
(line~\ref{alg:mon:start}), which ensures only robustness values needed for resulting robustness evaluation are computed.

\begin{algorithm}
	\caption{Robustness Monitoring for \STL{}}\label{alg:monitor}
	
\begin{algorithmic}[1]
	\Require \STL{} formula $\phi$ and timed state sequence $(\tau,\sigma)$ of length greater than $l(\phi)$ (see Definition~\ref{def:necessary_length}).
	\Ensure The value of $\rho(\phi,(\tau,\sigma))$.
	\Statex
	\State For any $i$ free in $\phi$, $\phi\gets\frz i\phi$.
		%Rewrite $\phi$, so that $\frin(\phi)=\emptyset$, using Theorem~\ref{thm:sentence}.
	%\State Minimize number of indices used in $\phi$.
		%Optimize $\phi$ using Algorithms~\ref{alg:movedown+merge} and \ref{alg:index_minimize} (respectively).
	\Statex
	\State$P\gets\emptyset$	\Comment{Precomupted robustness values.}
	\State\Return\Call{Monitor}{$\phi,0,\vc{0}$}	\label{alg:mon:start}
	\Statex

	\Procedure{Monitor}{$\phi,\iota,\iota*$}\label{alg:mon:monitor}
		\If{$\phi\equiv\top$}
			\Return$+\infty$
		\ElsIf{$\phi\equiv\mu$}
			\Return$\rho(\mu,(\tau,\sigma),\iota,\iota*)$	\Comment{According to Definition~\ref{def:predicate_robustness}.}
		\ElsIf{$\phi\equiv\neg\phi_1$}
			\Return$-\mbox{\Call{Monitor}{$\phi_1,\iota,\iota*$}}$
		\ElsIf{$\phi\equiv\phi_1\lor\phi_2$}
			\Return$\max\left(\mbox{\Call{Monitor}{$\phi_1,\iota,\iota*$}},\mbox{\Call{Monitor}{$\phi_2,\iota,\iota*$}}\right)$
		\ElsIf{$\phi\equiv\frz i\phi_1$}
			\Return\Call{Monitor}{$\phi_1,\iota,\store[\iota*]{i}{\iota}$}
		\ElsIf{$\phi\equiv\phi_1\U{a,b}\phi_2$}\label{alg:mon:until-beg}
			\If{$(\phi,\iota*)\in \dom(P)$}
				\State\Return$P(\phi,\iota*)(\iota)$
				%\State$\varrho\gets P(\phi,\iota*)$
				%\State\Return$\varrho_\iota$
			\Else
				\State$\varrho\gets\mbox{\Call{PrecomputeUntil}{$\phi_1,\phi_2,a,b,\iota*$}}$
				\State$P\gets P\cup((\phi,\iota*),\varrho)$
				\State\Return$\varrho_\iota$
			\EndIf\label{alg:mon:until-end}
		\EndIf
	\EndProcedure
	\Statex
	\Procedure{PrecomputeUntil}{$\phi_1,\phi_2,a,b,\iota*$}\label{alg:mon:until}
	\State$i\gets 0$
	\State$l\gets \max(l(\phi_1),l(\phi_2))$
	\State$\varrho\gets\emptyset$	\Comment{Sequence of robutness values.}
	\While{$\tau_i+b+l\le l(\tau)$}
		\State$j\gets 0$\label{alg:mon:out-beg}
		\State$r_1\gets \mbox{\Call{Monitor}{$\phi_1,i,\iota*$}}$
		\While{$\tau_{i+j}<\tau_i+a$}	\Comment{Before $[\tau_i+a,\tau_i+b]$.}
			\State$r_1\gets\min(r_1,\mbox{\Call{Monitor}{$\phi_1,i+j,\iota*$}})$
			\State$j\gets j+1$
		\EndWhile
		\State$r\gets r_1$
		\While{$\tau_{i+j}\le\tau_i+b$} \Comment{Inside $[\tau_i+a,\tau_i+b]$.}\label{alg:mon:in-beg}
			\State$r_1\gets\min(r_1,\mbox{\Call{Monitor}{$\phi_1,i+j,\iota*$}})$
			\State$r_2\gets\mbox{\Call{Monitor}{$\phi_2,i+j,\iota*$}}$
			\State$r\gets\max(r,\min(r_1,r_2))$
			\State$j\gets j+1$
		\EndWhile\label{alg:mon:in-end}
		\State$\varrho\gets\varrho\cup\{(i,r)\}$	\Comment{Set the value of $\varrho_i$.}
		\State$i\gets i+1$	\label{alg:mon:out-end}
	\EndWhile
	\EndProcedure
	%\algstore{monitor}
\end{algorithmic}

\end{algorithm}

Robustness values with respect to subformulae of input formula are not stored. Instead, they are computed every time procedure \procedure{Monitor} is called on a given subformula.
The reasoning behind this practise is the following: For the majority of formulae, the value of robustness for given $\iota$ and $\iota*$ is obtained by a~simple -- constant-time -- operation
on just a single value of robustness (or two in the case of $\lor$). Additionally, the robustness with respect to predicates can be computed in constant time. %as a~result of \sectionname~\ref{sec:5:mu}.

The only operator where robustness depends on robustness values over
an interval is the until operator (and by extension all derived temporal operators). Consequently, robustness values associated with until operators are stored. Furthermore, when \procedure{Monitor}($\phi_1\U*I\phi_2,\iota,\iota*$) is called for the first time,
robustness values with respect to $\phi_1\U*I\phi_2$ for $\iota*$ and all $\iota'$ are precomputed (see lines \ref{alg:mon:until-beg}--\ref{alg:mon:until-end}) by the procedure \procedure{PrecomputeUntil},
which constitutes an algorithmic version of robustness definition for until operator. These precomputed values are expected to be referred to later, since robustness computation is restricted to time interval
$[0,l(\phi)]$ which comprises all input values necessary to evaluate $\rho(\phi,(\tau,\sigma))$.

\subsection{Complexity}

Apparently, the most time-consuming task of Algorithm~\ref{alg:monitor} is the \procedure{PrecomputeUntil} procedure, which is quadratic to the number of states in the input
timed state sequence. In the worst case it is called for each $\iota*$. Therefore, the complexity of Algorithm~\ref{alg:monitor} is in $\mathcal{O}\left(|\phi|\cdot n^{2|\II|}\right)$ where
$n$ is the size of input timed state sequence. For sampled signals, it may be expressed using necessary length, resulting in alternate complexity formulation: $\mathcal{O}\left(|\phi|\cdot l(\phi)^{2|\II|}\cdot f^{2|\II|}\right)$
where $f$ is the sampling rate of input signal, which correlates with the precision of robustness computation. Space complexity can be bounded by the same function.

The parameter most adversely affecting the algorithm complexity is the size of frozen time index set~$|\II|$. Naturally, $\II$ can be restricted to indices used in input formula.
In most practical cases, their number will be small. This is supported by the following result:
\begin{theorem}
	Any formula $\phi$ can be rewritten into a~semantically equivalent formula which uses only so many indices as is the maximum number of free indices in subformulae of $\phi$.
\end{theorem}
Note that the number of free indices may increase as we descend into subformulae.

This statement derives from the fact that an index only serves to associate one freeze operator with a~set of coefficients in one or more predicates and it is free on all paths between
this freeze operator and all associated predicates. Therefore, indices which are never simultaneously free need not be different.

The result of this theorem can be realized by an automatic procedure which renames frozen time indices in a~formula while traversing its syntax tree (using DFS). This procedure stores pairs of indices $[k/l]$
corresponding to the renaming of source index $k$ in the original formula $\phi$ to destination index $l$ in its optimized version $\phi'$. When the procedure encounters freeze operator $\frz i$,
new pair $[i/m]$ is introduced where $m$ is the smallest unused destination index and the operator is changed to $\frz m$. Whenever $k$ becomes free in $\phi$, the pair $[k/l]$ is removed and $l$ can be reused.
Upon reaching a~predicate, all stored pairs are applied as a~renaming.
%% je tomu rozumět?

This procedure is described in greater detail in \cite{src} (page 44) where additional justification of its correctness can also be found.

Together with freeze operator merging described in Section~\ref{sec:merge_indices} (which does not increase number of free indices), this can considerably decrease the number of indices used in a~formula
and thus the time complexity of robustness monitoring. Although intelligent formula specification may result in already optimal formula, the existence of automatic optimization procedures
reduces demands on writers of formulae. %nějaký jiný termín? anyone? -- prostě jednodušší specifikace formulí

%TODO jak přesně se bude navazovat na tu optimalizaci
%, which is supported by the results of \sectionname~\ref{sec:4:minimize}.
%Introduction of even one frozen time index allows us to specify a~wide variety of biologically interesting properties as it has been argued by \citeneed{Dluhos-stlstar}.

\subsection{Implementation}

%\enlargethispage{5mm}

The algorithm has been implemented as an extension of the tool Parasim~\cite{parasim}. Parasim is a~highly modular Java-based open-source tool with graphical user interface for computing robustness of a~model with respect to perturbations. Integrating the algorithm presented in this paper into an already existing tool has an additional advantage of facilitating the use of \STL{} robustness in practise.

%In its function, it is similar to Breach~\cite{STL-breach}:
 Given a~model, \STL{} formula and perturbation set, Parasim samples the perturbation set into points and for each point simulates the model and computes robustness of the resulting signal with respect to \STL{} robustness measure. In the neighbourhood of signals with low robustness, additional points are sampled.
%One of the advantages of Parasim is its modular architecture which enables its efficient extension \cite{parasim}.
Formula optimizing algorithms are implemented to maximize efficiency.

%Regarding the employed technologies, Parasim is based on Java programming language \cite{java} and accepts input in the form of Extensible Markup Language documents \cite{XML}.
%The extension described in this section adopts these as well.

%Parasim and its user documentation are freely available on the Parasim website \cite{parasim}. 

% inzenyrsky prostup skrz vzorkovani

% od predikatu po operatory

% monitorujeme v diskretni semantics 

% algoritmus uvest

% rank minimization, nema smysl zabyhat do algoritmu, je technicky, jen rict co umime a kde je to uvedeno (DP)

% diskuse complexity

\section{Case Study}
\label{sec:casestudy}

By employing the Parasim tool we have conducted several experiments on two simple population dynamics models. The experiments have also served us to briefly evaluate the algorithm performance (in the setting of the Parasim tool).

%\enlargethispage{8mm}

\subsection{SIR Model}\label{sec:6:SIR}%{{{
\begin{figure}[h!]%{{{
\vspace*{-5mm}

\begin{center}
     \subfloat[]{
          \centering
          \includegraphics[scale=0.4]{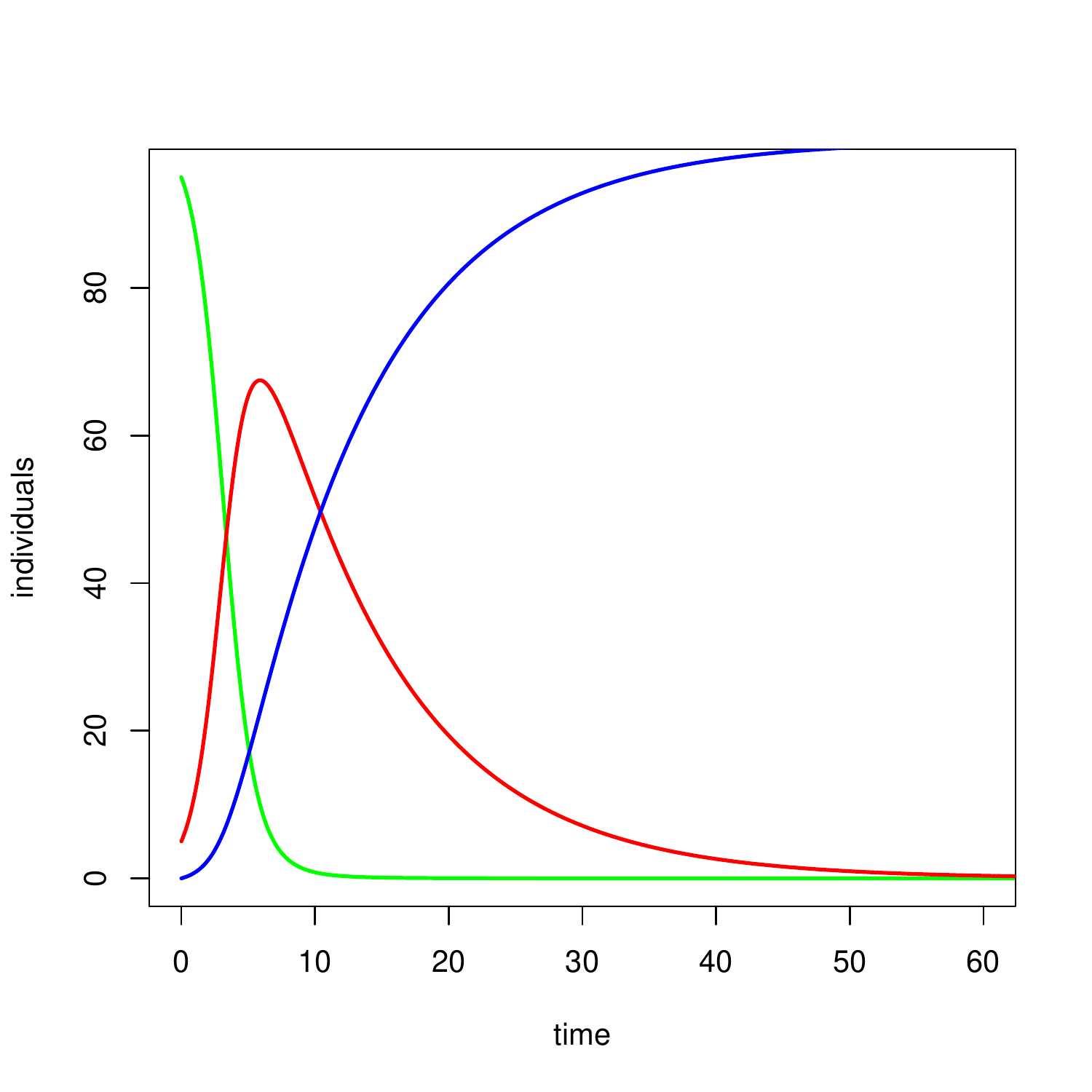}
      }
      \subfloat[]{
        \centering
        \includegraphics[scale=0.4]{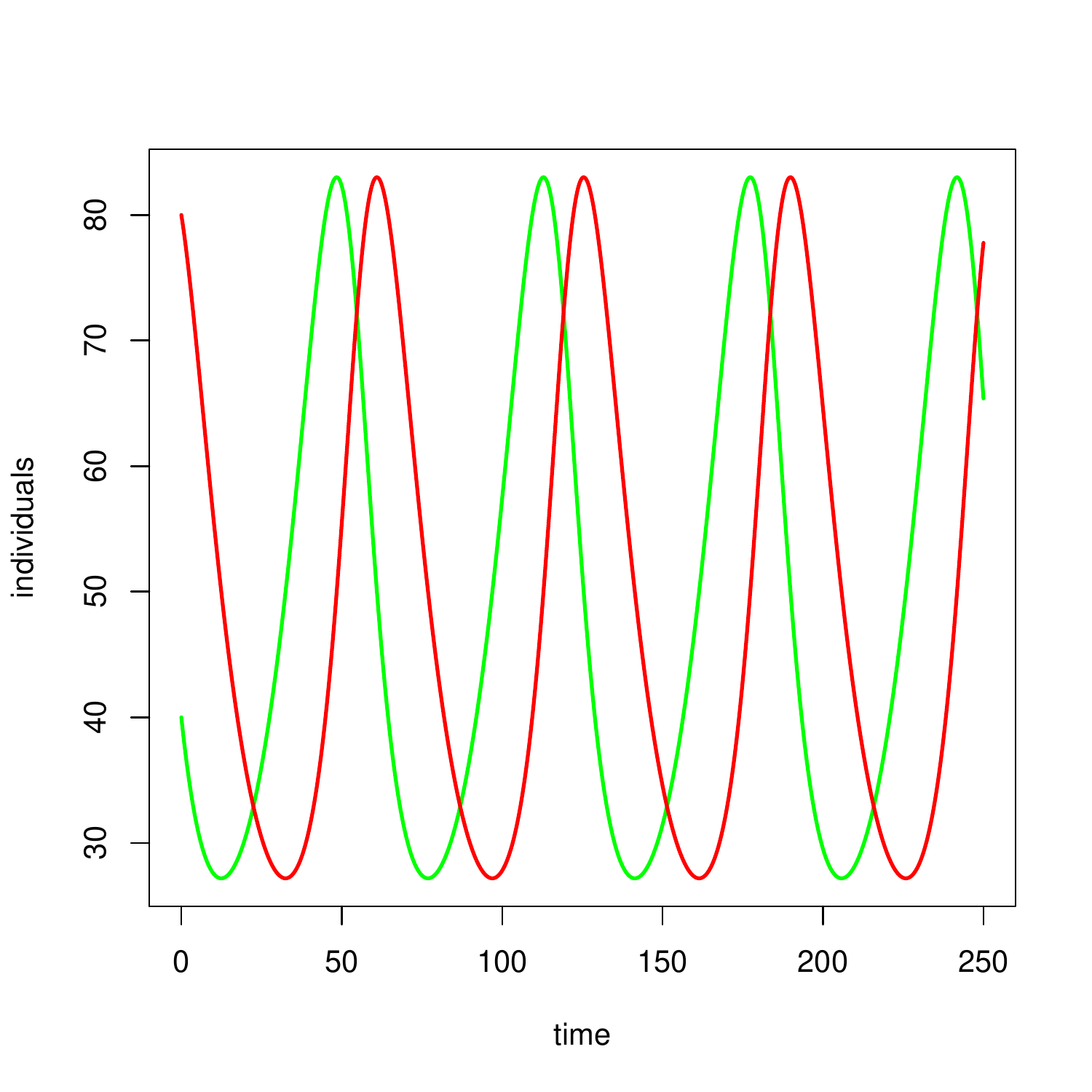}
      }
\vspace*{-4mm}
\end{center}
        \caption{(a) Typical development of SIR model, showing the number
                of susceptible (green), infected (red) and recovered (blue) individuals. (b) Typical development of populations in predator-prey model, showing number of prey (green) and predator (red).}\label{fig:sir-course}
\end{figure}

First, we demonstrate the robustness analysis on the model simulating an outbreak of an infectious disease in a~population~\cite{SIR}.
The simulated population is divided into three categories: \emph{susceptible} ($S$),
\emph{infected} ($I$) and \emph{recovered} ($R$). A~susceptible individual can become infected
by contact with another infected individual and an infected individual may recover. The ODE model is the following:
\begin{align*}
        \frac{dS}{dt}&=-\a SI &
        \frac{dI}{dt}&=\a SI-\beta I&
        \frac{dR}{dt}&=\beta I
\end{align*}
Where $\a$ is the \emph{contact rate} which correlates to probability of disease transmission, while $\beta$, the \emph{recovery rate},
takes into account the standard length of recovery. A~typical simulation of this model (see \figurename~\ref{fig:sir-course}a) includes
a~rapid increase in infected individuals, which is then followed by their gradual recovery.

%}}}

In this case study, we compare robustness analysis based on a formula containing value-freezing with respect to a freezing-free formula analysis exploiting a similar behavioural pattern. In particular, we consider the following formulae:
\begin{align*}
        \text{STL}:\phi_1&=\F{1,5}(I\ge 50) & 
        \text{STL*}:\phi_2&=\F{1,5}\left(I\ge 50\land\frz*\G{0.25,5}(I*\ge I)\right)
\end{align*}
Both formulae require the number of infected individuals to be greater than $50$ at some time in the interval $[1,5]$,
while $\phi_2$ also requires this number to be the local maximum (the number of infected individuals is required to decrease after reaching this maximum).

\begin{figure}[h!]%{{{
        \subfloat[Robustness wrt $\phi_1$.]{
                \centering
                \includegraphics[scale=.5]{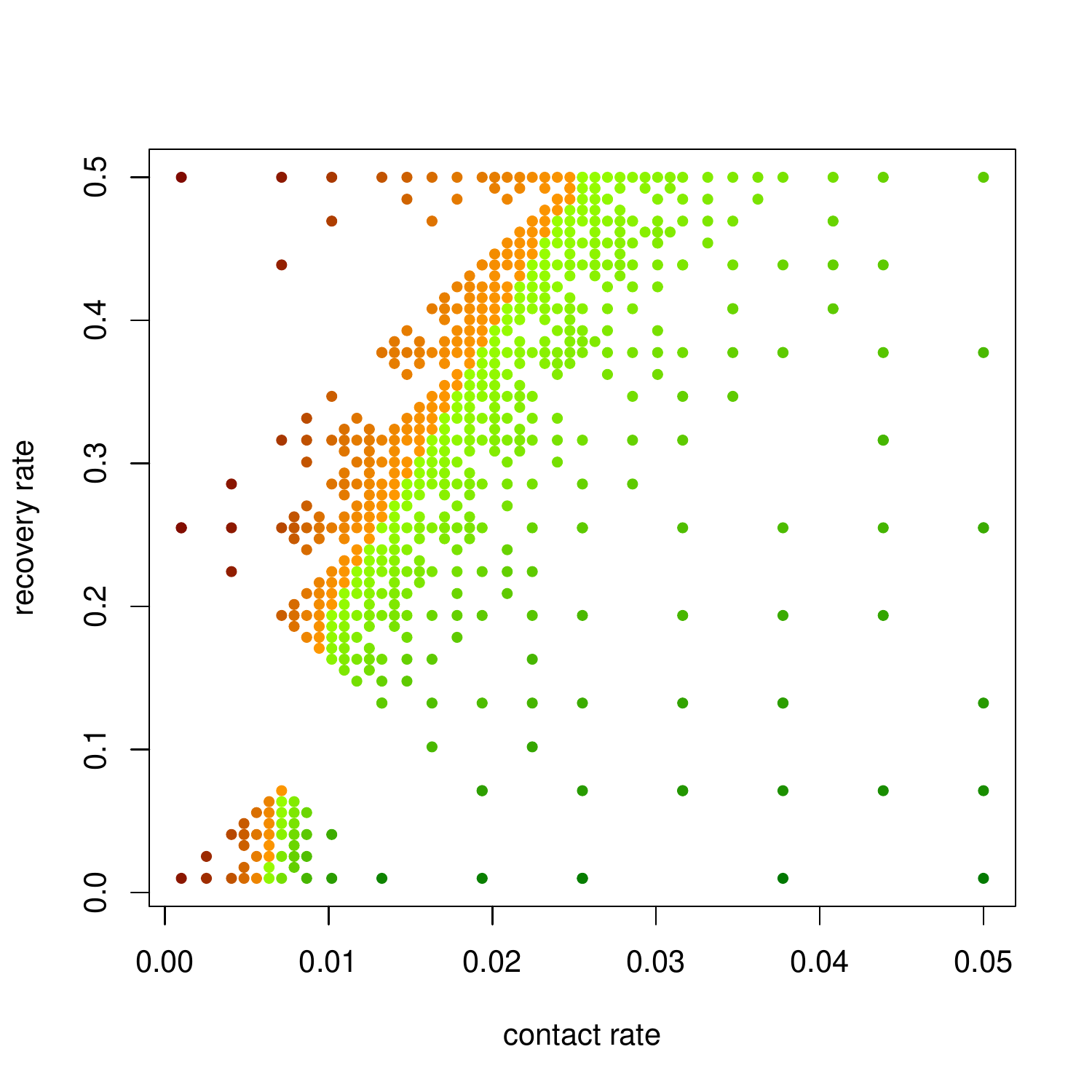}
                \label{fig:sir-out:normal}
        }
        \subfloat[Robustness wrt $\phi_2$.]{
                \centering
                \includegraphics[scale=.5]{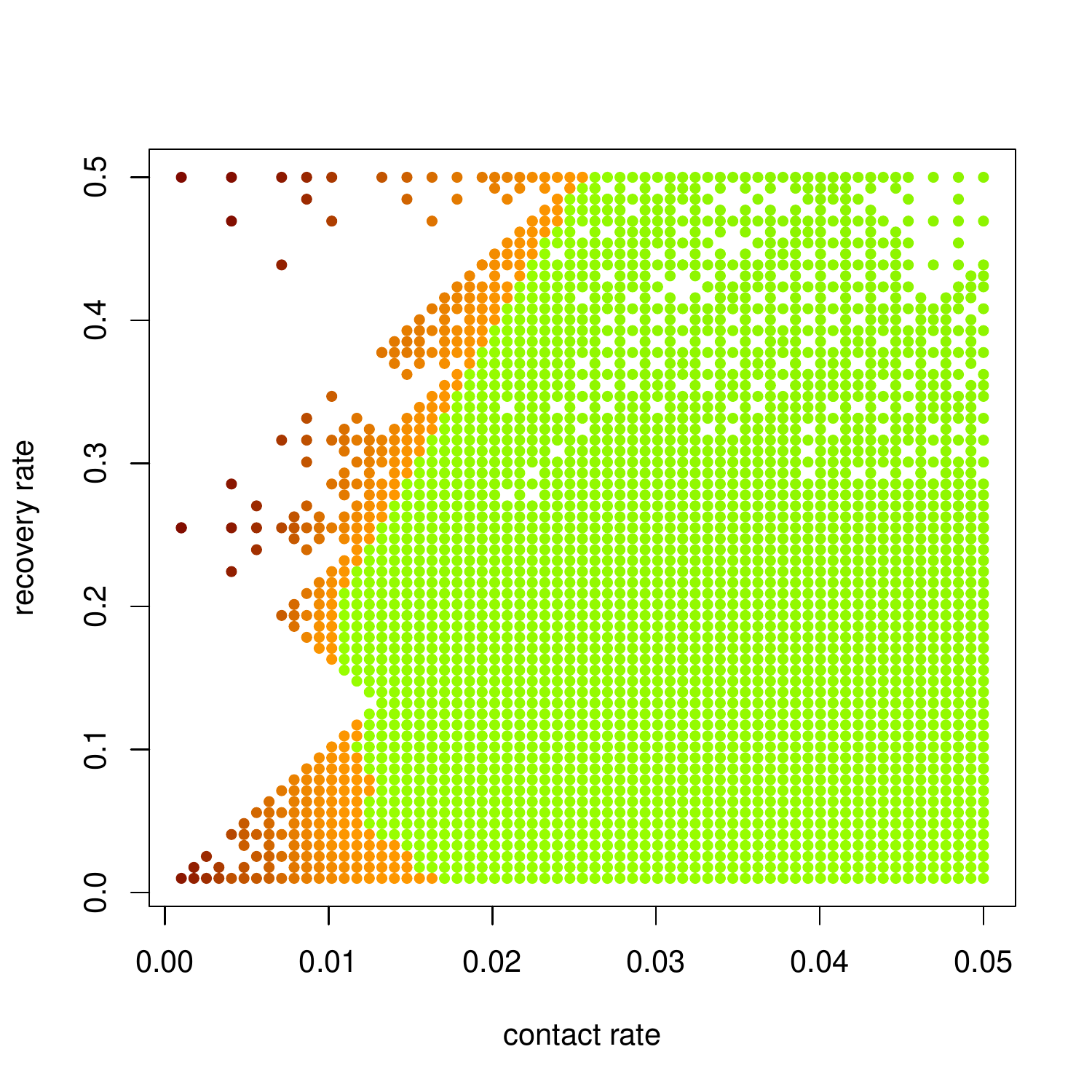}
                \label{fig:sir-out:star}
        }
        \caption[Robustness of SIR model with respect to contact and recovery rates.]{Robustness of SIR model with respect to $\phi_1$ and $\phi_2$ for variable contact and recovery rates. Robustness was positive in green points and negative in orange points. Darker colour represents greater absolute value of robustness.}\label{fig:sir-out}
\end{figure}

The robustness with respect to both properties was analysed on perturbations of both contact rate and recovery rate. Results are presented in \figurename~\ref{fig:sir-out}.

%}}}

While the satisfaction sets of $\phi_1$ and $\phi_2$ (delineated by positive robustness) are essentially identical, the actual robustness values show a~significant difference. Generally, when they are positive, the value of robustness
with respect to $\phi_1$ at given point is considerably greater than the corresponding value of robustness with respect to $\phi_2$.
In \figurename~\ref{fig:sir-out}, this can be seen as lighter shade of green points in \ref{fig:sir-out:star}. Also,
lower robustness causes the apparent increase in the number of points. %(see \sectionname~\ref{sec:6:impl}).
%D. uplne nerozumim "lighter shade", to musime trochu zlepsit ten popis

The reason for the rapid change in robustness comes from evaluation of the subformula $\frz*\G{0.25,5}(I*\ge I)$ that describes the local extreme. When evaluated in time $t$, robustness is proportional to the difference $(I[t]-I[t+0.25])$ (by Definition~\ref{def:predicate_robustness}). In practise, the difference is small provided that the descent of $I$ is not extremely steep. This causes such formulae to have typically low robustness values on common signals.

%second case study (in a shortened form, with a discussion put to conclusions)
\subsection{Predator-Prey Model}
\label{sec:6:lv}

In the second case study we analyse the predator-prey model \cite{lotka,volterra}, which attains oscillating behaviour for a~wide variety of parameters. We use a variant of the Lotka-Volterra model represented by the following ordinary differential equations:
\begin{align*}
        \frac{dX}{dt}&=\nu X-\a XY &
        \frac{dY}{dt}&=\a XY-\mu Y
\end{align*}
The model simulates a~situation where a~prey species~$X$ is hunted by a~predator 
species~$Y$ with the simplifying assumption that predator birth rate
and prey death rate are equal and proportional to the probability of prey and predator contact, and thus to the product of both species populations.
We use the following coefficients: prey natality ($\nu$), predator mortality ($\mu$) and predation rate ($\a$). Typical behaviour of this models
constitutes periodic oscillations (see \figurename~\ref{fig:sir-course}b).

We consider perturbation of two aforementioned coefficients, $\nu$ and $\a$, and compute robustness with respect to two properties specified by the following formulae:
{\small
\begin{align*}
        \psi_1 &=\G{0,300}\frz*\F{0,100}\left(X\ge Y*\right)\\
        \psi_2 &={\G{0,300}\left(X\ge 1\land Y\ge 1\land\F{0,50}\frz*\left(\F{0,75}\left(X*-X\ge 25\right)\land\F{0,75}\left(X-X*\ge 25\right)\right)\right)}
\end{align*}}
The property $\psi_1$ requires that for each time point $t\in[0,300]$, there is a~subsequent time point $t'\in[t,t+100]$ such that
population of prey in $t'$ is greater than population of predators in $t$. According to Definition~\ref{def:predicate_robustness} its corresponding robustness can be expressed as follows:
$$\rho(\phi,s)=\min_{t\in[0,300]}\max_{t'\in[t,t+100]}\frac{X[t']-Y[t]}{2}$$
where $X[t']$ and $Y[t]$ denote values of $s$ associated with given species at given time. The robustness value is maximized with respect to $t'$ and minimized with respect
to $t$, therefore, it uses maximal values of both $X$ and $Y$. Consequently, this property can be interpreted as maximum population of prey being greater then maximum population of predators (restricted to given intervals).

Formula $\psi_2$ is based on the similar principle. While rejecting aberrant behaviour where population of one of the species drops below one individual,
intuitively, it requires that there always is time in the future when population of prey can increase or decrease by 25 individuals, which is stated by the subformula
$\F{0,50}\frz*\left(\F{0,75}\left(X*-X\ge 25\right)\land\F{0,75}\left(X-X*\ge 25\right)\right)$.
Therefore, $\psi$ is satisfied when the difference between maximal and minimal prey population is greater than~50 and the associated robustness
is proportional to this difference. Again, we have avoided use of the extreme property, which would adversely affect robustness value.

%\enlargethispage{5mm}

\begin{figure}[h!]%{{{
\vspace*{-6mm}
\begin{center}
   %     \subfloat[Robustness wrt $\psi_1$.]{
                \includegraphics[scale=0.5]{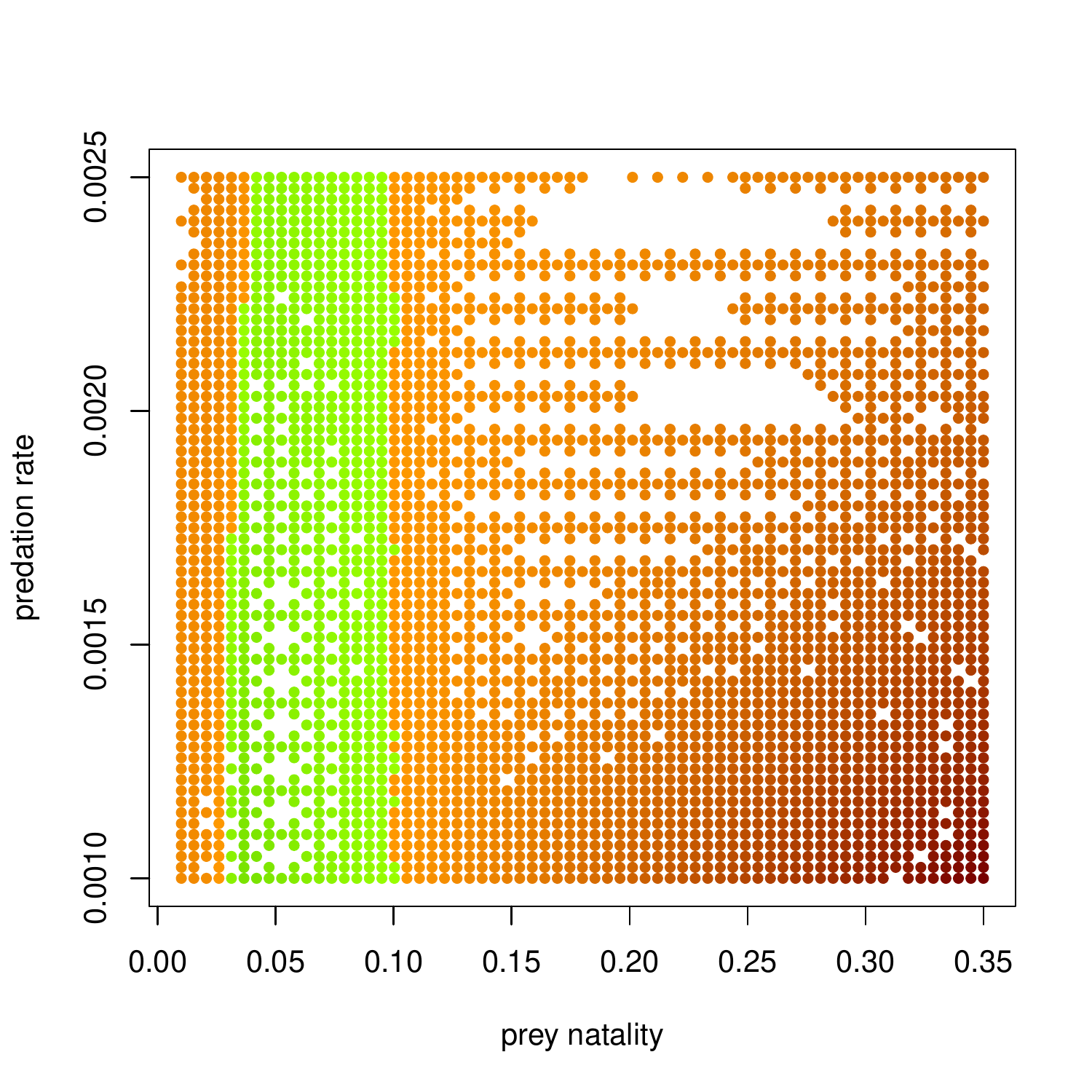}~~~
   %             \label{fig:lv-out:maxrel}
   %     }
   %     \subfloat[Robustness wrt $\psi_2$.]{
                \includegraphics[scale=0.5]{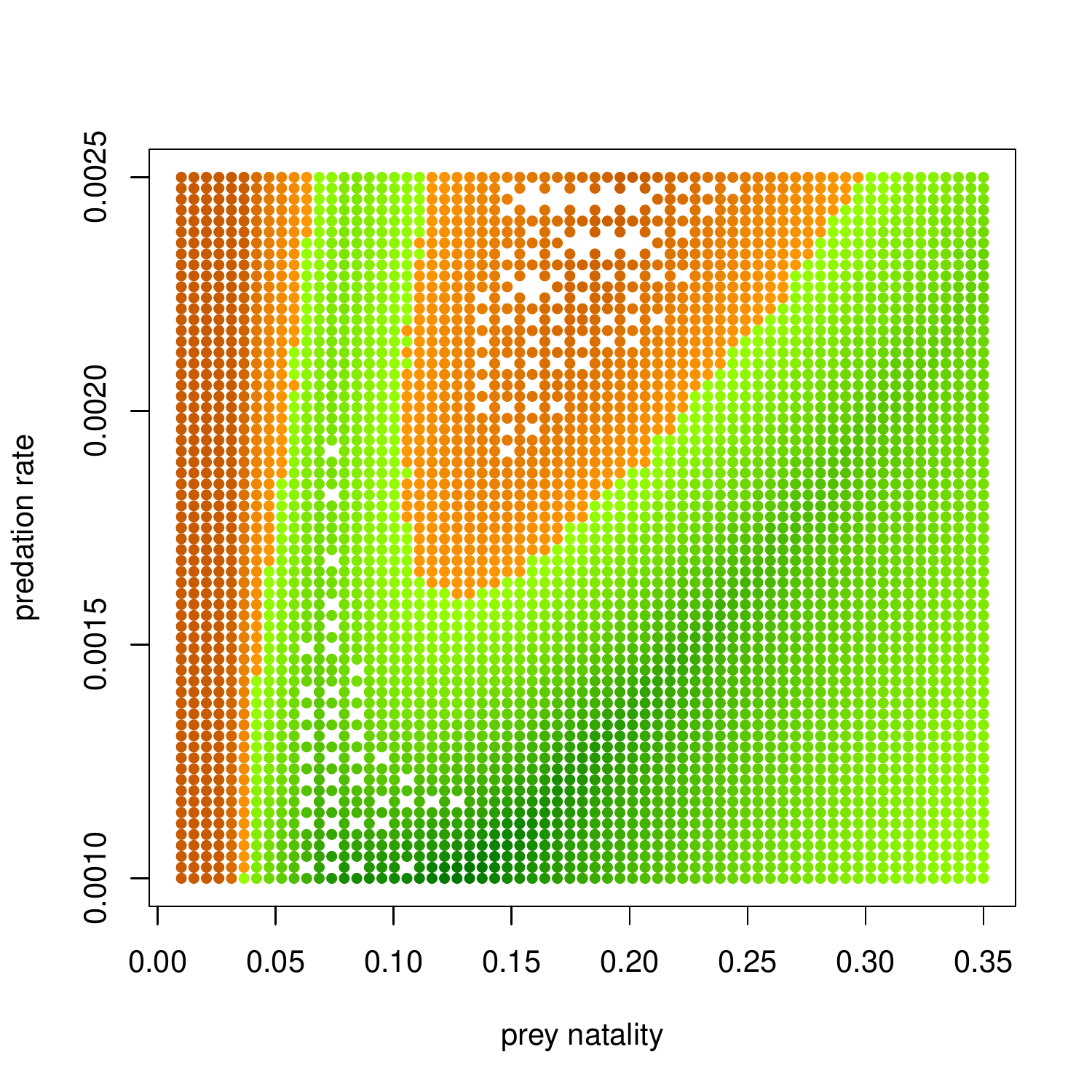}
   %             \label{fig:lv-out:oscil}
   %     }
\end{center}
        \caption[Robustness of predator-prey model with respect to prey natality and predation rate.]{Robustness of predator-prey model with respect to $\psi_1$ (left) and $\psi_2$ (right) for variable
        prey natality and predation rate. Robustness was positive in green points and negative in orange points. Darker colour represents greater absolute value of robustness.}\label{fig:lv-out}
\end{figure}%}}}

Results of this analysis are presented in \figurename~\ref{fig:lv-out}. Here, we should point out that small prey natality produced behaviour where
predator population approached zero and period of oscillations was greatly increased. For such behaviour, intervals used in $\psi_1$ and $\psi_2$ were shorter
than one period.

Apparently, satisfaction of $\psi_1$ is not affected by predation rate. More interestingly, when prey natality increases, predator population exceeds that of prey (see \figurename~\ref{fig:lv-out} (left)).
\figurename~\ref{fig:lv-out} (right) shows that amplitude of prey population oscillation is affected by both prey natality and predation rate.

The above results have been confirmed by simulation.

\subsection{Performance}

\enlargethispage{5mm}

Performance of robustness analysis is summarized in Table~\ref{tab:perf}. All results have been obtained by executing the algorithm implementation on a $4$ core $2$ GHz CPU with $4$ GB RAM. Each computation has been arranged into $8$ threads. For each analysis we have set an optimal resolution of the trajectories (number of simulated points). The number of simulated trajectories has been bounded by the number of refinement iterations in the Parasim parameter space sampling procedure. 

It is worth noting that all analysed properties consist only of $\F*{}$ and $\G*{}$ operators 
for which the procedure is optimized by employing Lemire queues in the same way as proposed in~\cite{DFM13}. This is based on an optimal streaming algorithm for computing maxima (resp. minima) of a numerical sequence and allows to reduce the quadratic complexity wrt formula size to linear.

\begin{table}[!h]
{\small
\begin{center}
\begin{tabular}{|c||c|c|c|c|}
\hline
Property (model) & formula size & $\#$ trajectories & $\#$ points per a trajectory & time\\
\hline\hline
$\varphi_1$ (SIR) & 2 & 250 & 500 & $8.6$ s\\
$\varphi_2$ (SIR) & 6 & 1365 & 1000 & $15.2$ s\\\hline
$\psi_1$ (Predator-Prey) & 4 & 831 & 400 & $85.4$ s\\
$\psi_2$ (Predator-Prey) & 12 & 1293 & 423 & $309.4$ s\\
\hline
\end{tabular}
\end{center}}
\caption{Performance of the robustness computation measured on the prototype implementation.}
\label{tab:perf}
\end{table}

The increase in computation time in the case of $\psi_1$ is caused by longer time intervals quantifying the temporal operators. Computation of the property $\psi_2$ has been slowed down due to insufficient memory.

% systemova biologie :-)

% zminit PARASIM

% case study I -- message ze to bezi a ze nektere vlastnosti delaji problem

% case study II -- message, ze lze merit robustnost oscilacniho chovani a problemem je jake volit intervaly, jak vymyslet vlastnosti, pridat jen vlastrnosti je case study I (v pripade nedostatku prostoru)

% dal bych nekam cca casy, v jake skale se to pohybuje, pro urc pocet bodu -- viz experimenty.txt

% taky je třeba zmínit, že v praxi se složitost odvíjí především od délky použitých intervalů (navíc pokud je před temporálním operátorem nějaká hvězdička)

\section{Conclusion}

In this paper we have set up a robustness measure for a value-freezing extension of STL. The robustness of a signal with respect to a given \STL{} property is based on the distance of the signal from signals violating the property. We have introduced a measure that is proved to fulfil requirements imposed on robustness measures as defined in~\cite{FP-robustness}. This guarantees that the robustness measure is defined correctly. We have derived the algorithm for \STL{} robustness computation from the discrete robustness and implemented it as an extension of the tool Parasim~\cite{parasim}.

Some of the properties from case studies required comparison of signal values at near frozen time points. Robustness of such properties is typically small.
This is only natural as such properties represent stricter requirements on signals. However, this feature may also constitute a~detriment for tools such as Parasim,
which use robustness to direct perturbation set sampling. This is the exact case of analysed SIR model and property $\phi_2$. It must be noted, though, that this problem is encompassed by the much broader issue of meaningful property design.

In~\cite{FP-robustness} the authors quantify error in robustness value caused by the approximate computation. We have not yet explored this possibility
for \STL{} robustness measures and leave this for future work. However, results in~\cite{FP-robustness} imply this error is inversely proportional
to the rate of input signal sampling.

\bibliographystyle{eptcs} % or whatever you prefer
\bibliography{hsb2013}

\begin{thebibliography}{10}
\providecommand{\bibitemdeclare}[2]{}
\providecommand{\surnamestart}{}
\providecommand{\surnameend}{}
\providecommand{\urlprefix}{Available at }
\providecommand{\url}[1]{\texttt{#1}}
\providecommand{\href}[2]{\texttt{#2}}
\providecommand{\urlalt}[2]{\href{#1}{#2}}
\providecommand{\doi}[1]{doi:\urlalt{http://dx.doi.org/#1}{#1}}
\providecommand{\bibinfo}[2]{#2}

\bibitemdeclare{article}{MITL}
\bibitem{MITL}
\bibinfo{author}{Rajeev \surnamestart Alur\surnameend},
  \bibinfo{author}{Tom\'{a}s \surnamestart Feder\surnameend} \&
  \bibinfo{author}{Thomas~A. \surnamestart Henzinger\surnameend}
  (\bibinfo{year}{1996}): \emph{\bibinfo{title}{{The Benefits of Relaxing
  Punctuality}}}.
\newblock {\sl \bibinfo{journal}{Journal of the ACM}}
  \bibinfo{volume}{43}(\bibinfo{number}{1}), pp. \bibinfo{pages}{116--146},
  \doi{10.1145/227595.227602}.

\bibitemdeclare{incollection}{FP-taliro}
\bibitem{FP-taliro}
\bibinfo{author}{Yashwanth Singh~Rahul \surnamestart Annapureddy\surnameend},
  \bibinfo{author}{Che \surnamestart Liu\surnameend},
  \bibinfo{author}{Georgios~E. \surnamestart Fainekos\surnameend} \&
  \bibinfo{author}{Sriram \surnamestart Sankaranarayanan\surnameend}
  (\bibinfo{year}{2011}): \emph{\bibinfo{title}{{S-TaLiRo: A~Tool for Temporal
  Logic Falsification for Hybrid Systems}}}.
\newblock In: {\sl \bibinfo{booktitle}{{Tools and Algorithms for the
  Construction and Analysis of Systems}}}, {\sl \bibinfo{series}{Lecture Notes
  in Computer Science}} \bibinfo{volume}{6605}, \bibinfo{publisher}{Springer},
  pp. \bibinfo{pages}{254--257}, \doi{10.1007/978-3-642-19835-9\_21}.

\bibitemdeclare{incollection}{SFM}
\bibitem{SFM}
\bibinfo{author}{Lubo\v{s} \surnamestart Brim\surnameend},
  \bibinfo{author}{Milan \surnamestart \v{C}e\v{s}ka\surnameend} \&
  \bibinfo{author}{David \surnamestart \v{S}afr\'anek\surnameend}
  (\bibinfo{year}{2013}): \emph{\bibinfo{title}{Model Checking of Biological
  Systems}}.
\newblock In \bibinfo{editor}{Marco \surnamestart Bernardo\surnameend},
  \bibinfo{editor}{Erik \surnamestart Vink\surnameend},
  \bibinfo{editor}{Alessandra \surnamestart Pierro\surnameend} \&
  \bibinfo{editor}{Herbert \surnamestart Wiklicky\surnameend}, editors: {\sl
  \bibinfo{booktitle}{Formal Methods for Dynamical Systems}}, {\sl
  \bibinfo{series}{Lecture Notes in Computer Science}} \bibinfo{volume}{7938},
  \bibinfo{publisher}{Springer Berlin Heidelberg}, pp.
  \bibinfo{pages}{63--112}, \doi{10.1007/978-3-642-38874-3\_3}.

\bibitemdeclare{article}{biocham}
\bibitem{biocham}
\bibinfo{author}{Laurence \surnamestart Calzone\surnameend},
  \bibinfo{author}{Fran\c{c}ois \surnamestart Fages\surnameend} \&
  \bibinfo{author}{Sylvain \surnamestart Soliman\surnameend}
  (\bibinfo{year}{2006}): \emph{\bibinfo{title}{{BIOCHAM: An Environment for
  Modeling Biological Systems and Formalizing Experimental Knowledge}}}.
\newblock {\sl \bibinfo{journal}{Bioinformatics}}
  \bibinfo{volume}{22}(\bibinfo{number}{14}), pp. \bibinfo{pages}{1805--1807},
  \doi{10.1093/bioinformatics/btl172}.

\bibitemdeclare{book}{modelchecking}
\bibitem{modelchecking}
\bibinfo{author}{Edmund~M. \surnamestart Clarke\surnameend},
  \bibinfo{author}{Orna \surnamestart Grumberg\surnameend} \&
  \bibinfo{author}{Doron~A. \surnamestart Peled\surnameend}
  (\bibinfo{year}{2000}): \emph{\bibinfo{title}{{Model Checking}}}.
\newblock \bibinfo{publisher}{MIT Press}.

\bibitemdeclare{book}{nonsmooth_analysis}
\bibitem{nonsmooth_analysis}
\bibinfo{author}{Frank~H. \surnamestart Clarke\surnameend}
  (\bibinfo{year}{1983}): \emph{\bibinfo{title}{{Optimization and Nonsmooth
  Analysis}}}.
\newblock \bibinfo{publisher}{S.I.A.M.}

\bibitemdeclare{inproceedings}{Dluhos-stlstar}
\bibitem{Dluhos-stlstar}
\bibinfo{author}{Petr \surnamestart Dluho\v{s}\surnameend},
  \bibinfo{author}{Lubo\v{s} \surnamestart Brim\surnameend} \&
  \bibinfo{author}{David \surnamestart \v{S}afr\'anek\surnameend}
  (\bibinfo{year}{2012}): \emph{\bibinfo{title}{{On Expressing and Monitoring
  Oscillatory Dynamics}}}.
\newblock In: {\sl \bibinfo{booktitle}{{Proceedings First International
  Workshop on Hybrid Systems and Biology}}}, \bibinfo{publisher}{Open Publ.
  Assoc.}, pp. \bibinfo{pages}{73--87}, \doi{10.4204/EPTCS.92.6}.

\bibitemdeclare{incollection}{STL-breach}
\bibitem{STL-breach}
\bibinfo{author}{Alexandre \surnamestart Donz\'e\surnameend}
  (\bibinfo{year}{2010}): \emph{\bibinfo{title}{Breach, A Toolbox for
  Verification and Parameter Synthesis of Hybrid Systems}}.
\newblock In \bibinfo{editor}{Tayssir \surnamestart Touili\surnameend},
  \bibinfo{editor}{Byron \surnamestart Cook\surnameend} \&
  \bibinfo{editor}{Paul \surnamestart Jackson\surnameend}, editors: {\sl
  \bibinfo{booktitle}{Computer Aided Verification}}, {\sl
  \bibinfo{series}{Lecture Notes in Computer Science}} \bibinfo{volume}{6174},
  \bibinfo{publisher}{Springer Berlin Heidelberg}, pp.
  \bibinfo{pages}{167--170}, \doi{10.1007/978-3-642-14295-6\_17}.

\bibitemdeclare{incollection}{STL-parameters}
\bibitem{STL-parameters}
\bibinfo{author}{Alexandre \surnamestart Donz\'{e}\surnameend},
  \bibinfo{author}{Gilles \surnamestart Clermont\surnameend},
  \bibinfo{author}{Axel \surnamestart Legay\surnameend} \&
  \bibinfo{author}{Christopher~J. \surnamestart Langmead\surnameend}
  (\bibinfo{year}{2009}): \emph{\bibinfo{title}{{Parameter Synthesis in
  Nonlinear Dynamical Systems: Application to Systems Biology}}}.
\newblock In: {\sl \bibinfo{booktitle}{{Research in Computational Molecular
  Biology}}}, {\sl \bibinfo{series}{Lecture Notes in Computer Science}}
  \bibinfo{volume}{5541}, \bibinfo{publisher}{Springer}, pp.
  \bibinfo{pages}{155--169}, \doi{10.1007/978-3-642-02008-7\_11}.

\bibitemdeclare{incollection}{DFM13}
\bibitem{DFM13}
\bibinfo{author}{Alexandre \surnamestart Donz\'e\surnameend},
  \bibinfo{author}{Thomas \surnamestart Ferrère\surnameend} \&
  \bibinfo{author}{Oded \surnamestart Maler\surnameend} (\bibinfo{year}{2013}):
  \emph{\bibinfo{title}{Efficient Robust Monitoring for STL}}.
\newblock In \bibinfo{editor}{Natasha \surnamestart Sharygina\surnameend} \&
  \bibinfo{editor}{Helmut \surnamestart Veith\surnameend}, editors: {\sl
  \bibinfo{booktitle}{Computer Aided Verification}}, {\sl
  \bibinfo{series}{Lecture Notes in Computer Science}} \bibinfo{volume}{8044},
  \bibinfo{publisher}{Springer Berlin Heidelberg}, pp.
  \bibinfo{pages}{264--279}, \doi{10.1007/978-3-642-39799-8\_19}.

\bibitemdeclare{inproceedings}{STL-robustness}
\bibitem{STL-robustness}
\bibinfo{author}{Alexandre \surnamestart Donz{\'e}\surnameend} \&
  \bibinfo{author}{Oded \surnamestart Maler\surnameend} (\bibinfo{year}{2010}):
  \emph{\bibinfo{title}{{Robust Satisfaction of Temporal Logic over Real-Valued
  Signals}}}.
\newblock In: {\sl \bibinfo{booktitle}{{FORMATS 2010}}},
  \bibinfo{publisher}{Springer}, pp. \bibinfo{pages}{92--106},
  \doi{10.1007/978-3-642-15297-9\_9}.

\bibitemdeclare{manual}{parasim}
\bibitem{parasim}
\bibinfo{organization}{Faculty of Informatics, Masaryk University}
  (\bibinfo{year}{2013}): \emph{\bibinfo{title}{{Parasim: Tool for Parallel
  Simulations and Verification}}}.
\newblock \urlprefix\url{https://github.com/sybila/parasim/wiki}.

\bibitemdeclare{incollection}{FP-robustness0}
\bibitem{FP-robustness0}
\bibinfo{author}{Georgios \surnamestart Fainekos\surnameend} \&
  \bibinfo{author}{George \surnamestart Pappas\surnameend}
  (\bibinfo{year}{2006}): \emph{\bibinfo{title}{Robustness of Temporal Logic
  Specifications}}.
\newblock In \bibinfo{editor}{Klaus \surnamestart Havelund\surnameend},
  \bibinfo{editor}{Manuel \surnamestart Nunez\surnameend},
  \bibinfo{editor}{Grigore \surnamestart Rosu\surnameend} \&
  \bibinfo{editor}{Burkhart \surnamestart Wolff\surnameend}, editors: {\sl
  \bibinfo{booktitle}{Formal Approaches to Software Testing and Runtime
  Verification}}, {\sl \bibinfo{series}{Lecture Notes in Computer Science}}
  \bibinfo{volume}{4262}, \bibinfo{publisher}{Springer Berlin Heidelberg}, pp.
  \bibinfo{pages}{178--192}, \doi{10.1007/11940197\_12}.

\bibitemdeclare{article}{FP-robustness}
\bibitem{FP-robustness}
\bibinfo{author}{Georgios~E. \surnamestart Fainekos\surnameend} \&
  \bibinfo{author}{George~J. \surnamestart Pappas\surnameend}
  (\bibinfo{year}{2009}): \emph{\bibinfo{title}{{Robustness of Temporal Logic
  Specifications For Continuous-Time Signals}}}.
\newblock {\sl \bibinfo{journal}{Theoretical Computer Science}}
  \bibinfo{volume}{410}(\bibinfo{number}{42}), pp. \bibinfo{pages}{4262--4291},
  \doi{10.1016/j.tcs.2009.06.021}.

\bibitemdeclare{article}{limitcycles}
\bibitem{limitcycles}
\bibinfo{author}{Leon \surnamestart Glass\surnameend} \&
  \bibinfo{author}{JoelS. \surnamestart Pasternack\surnameend}
  (\bibinfo{year}{1978}): \emph{\bibinfo{title}{Prediction of limit cycles in
  mathematical models of biological oscillations}}.
\newblock {\sl \bibinfo{journal}{Bulletin of Mathematical Biology}}
  \bibinfo{volume}{40}(\bibinfo{number}{1}), pp. \bibinfo{pages}{27--44},
  \doi{10.1007/BF02463128}.

\bibitemdeclare{incollection}{eziocav11}
\bibitem{eziocav11}
\bibinfo{author}{Radu \surnamestart Grosu\surnameend}, \bibinfo{author}{Gregory
  \surnamestart Batt\surnameend}, \bibinfo{author}{FlavioH. \surnamestart
  Fenton\surnameend}, \bibinfo{author}{James \surnamestart Glimm\surnameend},
  \bibinfo{author}{Colas \surnamestart Guernic\surnameend},
  \bibinfo{author}{ScottA. \surnamestart Smolka\surnameend} \&
  \bibinfo{author}{Ezio \surnamestart Bartocci\surnameend}
  (\bibinfo{year}{2011}): \emph{\bibinfo{title}{From Cardiac Cells to Genetic
  Regulatory Networks}}.
\newblock In \bibinfo{editor}{Ganesh \surnamestart Gopalakrishnan\surnameend}
  \& \bibinfo{editor}{Shaz \surnamestart Qadeer\surnameend}, editors: {\sl
  \bibinfo{booktitle}{Computer Aided Verification}}, {\sl
  \bibinfo{series}{Lecture Notes in Computer Science}} \bibinfo{volume}{6806},
  \bibinfo{publisher}{Springer Berlin Heidelberg}, pp.
  \bibinfo{pages}{396--411}, \doi{10.1007/978-3-642-22110-1\_31}.

\bibitemdeclare{article}{sybi-periodic}
\bibitem{sybi-periodic}
\bibinfo{author}{Benno \surnamestart Hess\surnameend} (\bibinfo{year}{2000}):
  \emph{\bibinfo{title}{{Periodic Patterns in Biology}}}.
\newblock {\sl \bibinfo{journal}{Naturwissenschaften}}
  \bibinfo{volume}{87}(\bibinfo{number}{5}), pp. \bibinfo{pages}{199--211},
  \doi{10.1007/s001140050704}.

\bibitemdeclare{article}{kaplanthe2008}
\bibitem{kaplanthe2008}
\bibinfo{author}{Shai \surnamestart Kaplan\surnameend}, \bibinfo{author}{Anat
  \surnamestart Bren\surnameend}, \bibinfo{author}{Erez \surnamestart
  Dekel\surnameend} \& \bibinfo{author}{Uri \surnamestart Alon\surnameend}
  (\bibinfo{year}{2008}): \emph{\bibinfo{title}{The incoherent feed-forward
  loop can generate non-monotonic input functions for genes}}.
\newblock {\sl \bibinfo{journal}{Molecular Systems Biology}}
  \bibinfo{volume}{4}(\bibinfo{number}{1}), pp.~\bibinfo{pages}{--},
  \doi{10.1038/msb.2008.43}.

\bibitemdeclare{article}{SIR}
\bibitem{SIR}
\bibinfo{author}{William~O. \surnamestart Kermack\surnameend} \&
  \bibinfo{author}{Anderson~G. \surnamestart McKendrick\surnameend}
  (\bibinfo{year}{1927}): \emph{\bibinfo{title}{{A~Contribution to the
  Mathematical Theory of Epidemics}}}.
\newblock {\sl \bibinfo{journal}{Proceedings of the Royal Society of London.
  Series~A}} \bibinfo{volume}{115}(\bibinfo{number}{772}), pp.
  \bibinfo{pages}{700--721}, \doi{10.1098/rspa.1927.0118}.

\bibitemdeclare{article}{Kitano-robustness}
\bibitem{Kitano-robustness}
\bibinfo{author}{Hiroaki \surnamestart Kitano\surnameend}
  (\bibinfo{year}{2004}): \emph{\bibinfo{title}{{Biological Robustness}}}.
\newblock {\sl \bibinfo{journal}{Nature Reviews Genetics}}
  \bibinfo{volume}{5}(\bibinfo{number}{11}), pp. \bibinfo{pages}{826--837},
  \doi{10.1038/nrg1471}.

\bibitemdeclare{article}{MTL}
\bibitem{MTL}
\bibinfo{author}{Ron \surnamestart Koymans\surnameend} (\bibinfo{year}{1990}):
  \emph{\bibinfo{title}{{Specifying Real-Time Properties with Metric Temporal
  Logic}}}.
\newblock {\sl \bibinfo{journal}{Real-Time Systems}} \bibinfo{volume}{2}, pp.
  \bibinfo{pages}{255--299}, \doi{10.1007/BF01995674}.

\bibitemdeclare{article}{Krejci2004152}
\bibitem{Krejci2004152}
\bibinfo{author}{Pavel \surnamestart Krejci\surnameend},
  \bibinfo{author}{Vitezslav \surnamestart Bryja\surnameend},
  \bibinfo{author}{Jiri \surnamestart Pachernik\surnameend},
  \bibinfo{author}{Ales \surnamestart Hampl\surnameend},
  \bibinfo{author}{Robert \surnamestart Pogue\surnameend},
  \bibinfo{author}{Pertchoui \surnamestart Mekikian\surnameend} \&
  \bibinfo{author}{William~R \surnamestart Wilcox\surnameend}
  (\bibinfo{year}{2004}): \emph{\bibinfo{title}{{FGF2} inhibits proliferation
  and alters the cartilage-like phenotype of {RCS} cells}}.
\newblock {\sl \bibinfo{journal}{Experimental Cell Research}}
  \bibinfo{volume}{297}(\bibinfo{number}{1}), pp. \bibinfo{pages}{152 -- 164},
  \doi{10.1016/j.yexcr.2004.03.011}.

\bibitemdeclare{book}{lotka}
\bibitem{lotka}
\bibinfo{author}{Alfred~J. \surnamestart Lotka\surnameend}
  (\bibinfo{year}{1925}): \emph{\bibinfo{title}{Elements of Physical Biology}}.
\newblock \bibinfo{publisher}{Williams and Wilkins},
  \bibinfo{address}{Baltimore}.

\bibitemdeclare{incollection}{Maler_STL}
\bibitem{Maler_STL}
\bibinfo{author}{Oded \surnamestart Maler\surnameend} \& \bibinfo{author}{Dejan
  \surnamestart Nickovic\surnameend} (\bibinfo{year}{2004}):
  \emph{\bibinfo{title}{Monitoring Temporal Properties of Continuous Signals}}.
\newblock In \bibinfo{editor}{Yassine \surnamestart Lakhnech\surnameend} \&
  \bibinfo{editor}{Sergio \surnamestart Yovine\surnameend}, editors: {\sl
  \bibinfo{booktitle}{Formal Techniques, Modelling and Analysis of Timed and
  Fault-Tolerant Systems}}, {\sl \bibinfo{series}{Lecture Notes in Computer
  Science}} \bibinfo{volume}{3253}, \bibinfo{publisher}{Springer Berlin
  Heidelberg}, pp. \bibinfo{pages}{152--166},
  \doi{10.1007/978-3-540-30206-3\_12}.

\bibitemdeclare{article}{biocham-practice}
\bibitem{biocham-practice}
\bibinfo{author}{Elisabetta~De \surnamestart Maria\surnameend},
  \bibinfo{author}{Fran\c{c}ois \surnamestart Fages\surnameend},
  \bibinfo{author}{Aur\'{e}lien \surnamestart Rizk\surnameend} \&
  \bibinfo{author}{Sylvain \surnamestart Soliman\surnameend}
  (\bibinfo{year}{2011}): \emph{\bibinfo{title}{{Design, Optimization and
  Predictions of a~Coupled Model of the Cell Cycle, Circadian Clock, DNA Repair
  System, Irinotecan Metabolism and Exposure Control under Temporal Logic
  Constraints}}}.
\newblock {\sl \bibinfo{journal}{Theoretical Computer Science}}
  \bibinfo{volume}{412}(\bibinfo{number}{21}), pp. \bibinfo{pages}{2108--2127},
  \doi{10.1016/j.tcs.2010.10.036}.

\bibitemdeclare{article}{robustness-property}
\bibitem{robustness-property}
\bibinfo{author}{Aur{\'e}lien \surnamestart Rizk\surnameend},
  \bibinfo{author}{Gr{\'e}gory \surnamestart Batt\surnameend},
  \bibinfo{author}{Fran\c{c}ois \surnamestart Fages\surnameend} \&
  \bibinfo{author}{Sylvain \surnamestart Soliman\surnameend}
  (\bibinfo{year}{2011}): \emph{\bibinfo{title}{Continuous valuations of
  temporal logic specifications with applications to parameter optimization and
  robustness measures}}.
\newblock {\sl \bibinfo{journal}{Theor. Comput. Sci.}}
  \bibinfo{volume}{412}(\bibinfo{number}{26}), pp. \bibinfo{pages}{2827--2839},
  \doi{10.1016/j.tcs.2010.05.008}.

\bibitemdeclare{article}{FagesRobustness}
\bibitem{FagesRobustness}
\bibinfo{author}{Aurélien \surnamestart Rizk\surnameend},
  \bibinfo{author}{Gregory \surnamestart Batt\surnameend},
  \bibinfo{author}{François \surnamestart Fages\surnameend} \&
  \bibinfo{author}{Sylvain \surnamestart Soliman\surnameend}
  (\bibinfo{year}{2009}): \emph{\bibinfo{title}{A general computational method
  for robustness analysis with applications to synthetic gene networks}}.
\newblock {\sl \bibinfo{journal}{Bioinformatics}}
  \bibinfo{volume}{25}(\bibinfo{number}{12}), pp. \bibinfo{pages}{i169--i178},
  \doi{10.1093/bioinformatics/btp200}.

\bibitemdeclare{article}{STL-use}
\bibitem{STL-use}
\bibinfo{author}{\surnamestart {Szymon Stoma et al.}\surnameend}
  (\bibinfo{year}{2013}): \emph{\bibinfo{title}{STL-based Analysis of
  TRAIL-induced Apoptosis Challenges the Notion of Type I/Type II Cell Line
  Classification}}.
\newblock {\sl \bibinfo{journal}{PLoS Comput Biol}}
  \bibinfo{volume}{9}(\bibinfo{number}{5}), p. \bibinfo{pages}{e1003056},
  \doi{10.1371/journal.pcbi.1003056}.

\bibitemdeclare{mastersthesis}{src}
\bibitem{src}
\bibinfo{author}{Tom\'{a}\v{s} \surnamestart Vejpustek\surnameend}
  (\bibinfo{year}{2013}): \emph{\bibinfo{title}{{Robustness Analysis of
  Extended Signal Temporal Logic STL*}}}.
\newblock Master's thesis, \bibinfo{school}{Masaryk University, Faculty of
  Informatics}.
\newblock \urlprefix\url{http://is.muni.cz/th/324713/fi_m/}.

\bibitemdeclare{article}{volterra}
\bibitem{volterra}
\bibinfo{author}{Vito \surnamestart Volterra\surnameend}
  (\bibinfo{year}{1928}): \emph{\bibinfo{title}{Variations and Fluctuations of
  the Number of Individuals in Animal Species living together}}.
\newblock {\sl \bibinfo{journal}{Journal du Conseil}}
  \bibinfo{volume}{3}(\bibinfo{number}{1}), pp. \bibinfo{pages}{3--51},
  \doi{10.1093/icesjms/3.1.3}.

\end{thebibliography}

%\newpage
%\begin{appendix}
%	\begin{algorithm}
%		\input{alg-movedown-merge}
%	\end{algorithm}
%	\begin{algorithm}
%		\input{alg-minimize}
%	\end{algorithm}
%\end{appendix}

%\newpage

% \begin{appendix}
% \input{6_appendix_compmod.tex}
% \end{appendix}

\end{document}